\PassOptionsToPackage{table}{xcolor}
\documentclass[10pt]{article}

\usepackage{microtype}
\usepackage{graphicx}
\usepackage{subfigure}
\usepackage{booktabs} 
\usepackage{float}
\usepackage[T1]{fontenc}
\usepackage{courier}
\usepackage{listings}
\usepackage{enumitem}
\usepackage{needspace}
\usepackage[normalem]{ulem}

\usepackage[most]{tcolorbox}

\usepackage{hyperref}
\usepackage{multicol}
\usepackage{multirow}
\usepackage{wrapfig}

\definecolor{best}{HTML}{C6E0B4}
\definecolor{secondbest}{HTML}{E2F0D9}
\definecolor{worst}{HTML}{F4C2C2}
\definecolor{secondworst}{HTML}{F9E0E0}

\definecolor{best1}{HTML}{B2DFDB}
\definecolor{best2}{HTML}{CDEAE8}
\definecolor{best3}{HTML}{E0F2F1}
\definecolor{best4}{HTML}{F1F8F7}
\definecolor{worst1}{HTML}{FFCDD2}
\definecolor{worst2}{HTML}{F8E1E3}
\definecolor{worst3}{HTML}{FCEBED}
\definecolor{worst4}{HTML}{FFF8F9}

\tcbset{
  colback=best3,     
  colframe=best1,    
  boxrule=0.4pt,
  arc=2pt,
  left=6pt,right=6pt,top=5pt,bottom=5pt,
}

\newtcolorbox{promptbox}[1][]{
  enhanced,
  breakable,
  #1
}

\usepackage{iclr2027_conference,times}
\iclrfinalcopy

\usepackage[utf8]{inputenc} 
\usepackage[T1]{fontenc}    
\usepackage{hyperref}       
\usepackage{url}            
\usepackage{booktabs}       
\usepackage{amsfonts}       
\usepackage{nicefrac}       
\usepackage{microtype}      

\usepackage{amsmath}
\usepackage{amssymb}
\usepackage{mathtools}
\usepackage{amsthm}

\usepackage[capitalize,noabbrev]{cleveref}
\usepackage{tabularx} 
\theoremstyle{plain}

\theoremstyle{definition}

\theoremstyle{remark}

\usepackage[disable,textsize=tiny]{todonotes}

\colorlet{pendingblue}{black}
\DeclareRobustCommand{\revision}[1]{#1}
\DeclareRobustCommand{\newrevision}[1]{#1}
\DeclareRobustCommand{\deletedrevision}[1]{}
\newtcolorbox{experimentgap}[1]{enhanced,breakable,colback=white,colframe=pendingblue,coltext=pendingblue,colbacktitle=white,coltitle=pendingblue,fonttitle=\bfseries,title={#1 --- planned; results pending},boxrule=0.5pt,arc=1pt,left=6pt,right=6pt,top=5pt,bottom=5pt}

\title{Validated Behavioral Hypotheses as a Lens for Evaluating Participant Simulation}
\newcommand{\aff}[1]{{\normalfont\normalsize #1}}
\author{%
\makebox[0.95\textwidth][c]{%
\begin{tabular}[t]{@{}*{4}{>{\centering\arraybackslash}p{0.23\textwidth}}@{}}
Xuan Liu$^{1}$ & HaoYang Shang$^{2}$ & Zizhang Liu$^{3}$ & Yuanjun Feng$^{4}$ \\[0.5ex]
Guankai Zhai$^{5}$ & Yunze Xiao$^{6}$ & Yiwen Tu$^{1}$ & Haojian Jin$^{1}$ \\[1.0ex]
\multicolumn{4}{c}{\aff{$^{1}$University of California, San Diego \qquad $^{2}$Independent Researcher}} \\[0.2ex]
\multicolumn{4}{c}{\aff{$^{3}$Tsinghua University \qquad $^{4}$Universit\'e de Lausanne}} \\[0.2ex]
\multicolumn{4}{c}{\aff{$^{5}$Stanford University \qquad $^{6}$Carnegie Mellon University}} \\[0.6ex]
\multicolumn{4}{c}{\aff{\texttt{xul049@ucsd.edu}}}
\end{tabular}%
}%
}

\begin{document}
\maketitle
\lhead{Preprint}
\vspace{-0.8em}
\begin{figure}[H]
\centering
\color{pendingblue}
\includegraphics[width=\linewidth]{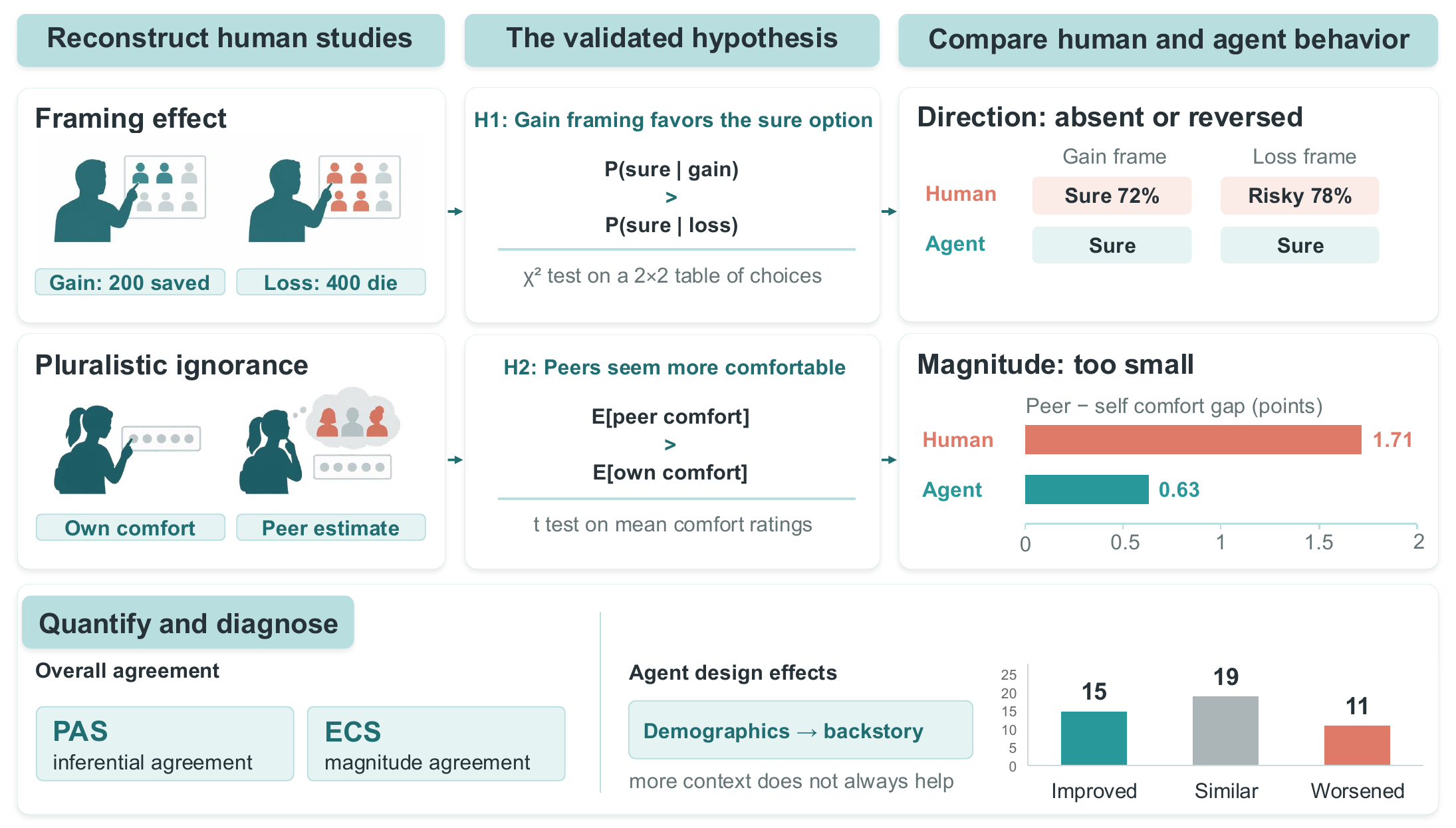}
\caption{\newrevision{\textbf{Behavioral hypotheses reveal where AI agents match and diverge from humans.} \textsc{HumanStudy-Bench} reconstructs human studies, administers them to agents role-playing as silicon participants, and tests whether agents reproduce the selected behavioral hypotheses. PAS measures inferential agreement; ECS measures effect-magnitude agreement. Top: \emph{framing effect} and \emph{pluralistic ignorance} show discrepancies in effect direction and magnitude. Bottom: agent design improves agreement with some hypotheses while worsening agreement with others.}}
\label{fig:overview}
\end{figure}
\begin{abstract}
{We propose using validated behavioral hypotheses as a lens for evaluating LLM agents as simulated human participants. This approach makes behavioral agreement in participant simulation measurable and decomposable, revealing which human effects agents reproduce, where they diverge, and how agent design changes that agreement. To operationalize this idea, we build \textsc{HumanStudy-Bench}, an open benchmarking platform that reconstructs experimental protocols from published human studies and administers these protocols to agents serving as \emph{silicon participants}. The benchmark compares population-level effects derived from agent responses with the corresponding published human findings using two metrics: the \emph{Probability Alignment Score} (PAS) for inferential agreement and the \emph{Effect Consistency Score} (ECS) for effect-magnitude agreement. We apply \textsc{HumanStudy-Bench} across 12 human studies, evaluating 10 models under four agent designs, with 6,588 simulated participants per agent configuration. By making published human studies and their validated behavioral hypotheses reusable for evaluating simulated participants, we hope to make the behavioral assumptions underlying LLM-based social simulations and their sensitivity to agent design explicit and empirically testable.}
\end{abstract}
\section{Introduction}
\label{sec:intro}

Large language models (LLMs) are being used as surrogate participants in social science experiments~\citep{anthis2025position,aher2023using,dillion2023can,gao2024large}, as testbeds for simulating user populations~\citep{moon2024virtualpersonaslanguagemodels,xiaohongshu_socialworldmodel,mou2024individualsocietysurveysocial}, and as building blocks of multi-agent simulations of organizations and markets~\citep{argyle2023out, Hofmann2024_nature, li2024frontiers, manning2024automated}. \revision{In these settings, models draw on their knowledge of human behavior to play human roles~\citep{shanahan2023roleplay}. These applications implicitly assume that agents reproduce relevant aspects of the behavior of the people they represent.} Yet evidence is accumulating that current LLM-based agents \revision{fall short of this assumption} in ways that compromise the validity of the simulations they ground~\citep{doi:10.1073/pnas.2501660122,participants_nature}.
\revision{Evaluating agents for these applications therefore requires a principled and diagnostic account of behavioral correspondence: we need to identify which human behavioral patterns agents reproduce, where they diverge, and how their configuration changes those patterns~\citep{puelmatouzel2026validationgap}.}

The conventional operationalization of the human-likeness test is Turing's \emph{imitation game}~\citep{turing1950computing}, in which a human judge interacts with a respondent and decides whether the respondent is human.
However, this approach has three key limitations. First, the verdict is subjective and sensitive to the judge's expectations of surface-level fluency~\citep{tjuatja2024llmsexhibithumanlikeresponse,xu2025languagemodelsmirrorhuman}.
Second, the outcome is a coarse binary that says nothing about \emph{which} aspects of human-likeness are missing.
Third, the protocol is expensive, requiring fresh human judges for every new agent. Indeed, as LLMs grow more sophisticated, passing such tests has increasingly become a question of stylistic mimicry rather than evidence of any deeper behavioral correspondence with humans~\citep{Gui_2023,Kitadai_2025replaceeco,grossmann2023ai}.

Decades of psychology, behavioral economics, and social psychology have produced testable claims about how human populations behave in experimental settings. We use subsequent independent human evidence to screen these behavioral hypotheses at two levels: \emph{direct} support tests the same or a closely matched contrast, whereas \emph{related} support establishes the broader behavioral effect. These hypotheses make quantitative predictions about effects such as framing~\citep{Tversky1981TheFO,kuhberger1998influence}, false consensus~\citep{Ross1977TheC,mullen1985false}, trust-game cooperation~\citep{Berg1995TrustRA,johnson2011trust}, and Asch impression formation~\citep{asch1946forming,nauts2014forming}. \revision{The human studies behind these hypotheses specify the tasks, conditions, and reported effects. Reconstructing those tasks and conditions for agents lets us compare population-level effects derived from their choices and judgments with the human findings, hypothesis by hypothesis. Screening hypotheses against follow-up evidence thus turns an abstract question about human-likeness into a falsifiable one: under corresponding experimental conditions, do agent populations reproduce the selected human effects?}

\revision{This hypothesis-by-hypothesis comparison reframes human-likeness evaluation from a single holistic judgment into a set of statistical comparisons against behavioral findings screened by follow-up evidence, offering three advantages over the imitation game.} \emph{Objective:} the assessment follows a fixed statistical pipeline applied to agents' responses, not a human judge's impression. \emph{Decomposable:} agreement and disagreement can be localized to specific hypotheses and dimensions of each effect. \emph{Scalable:} once an experimental protocol is reconstructed, any new agent can be administered the test end-to-end at low marginal cost.

To operationalize this idea, we build \textsc{HumanStudy-Bench}, an open platform that turns published human studies into reusable simulation environments.
\textsc{HumanStudy-Bench} introduces a human-in-the-loop pipeline that ingests published articles, extracts each study's hypotheses, experimental protocol, and statistical analysis methods, and compiles them into machine-executable experiment protocols for AI agents.
\newrevision{Working with domain experts, we curated an initial suite of 12 studies using independent evidence to guide effect selection (Table~\ref{tab:hypothesis-teaser} shows examples; Appendix~\ref{app:study} lists the suite).} 
\revision{The platform evaluates how well configured agents reproduce reported human behavioral effects under reconstructed experimental protocols.}

Given a set of AI agents, we run them through the reconstructed protocols as \emph{silicon participants} and test whether the resulting population-level effects agree with the selected human findings. \textsc{HumanStudy-Bench} uses two metrics to quantify this agreement.
The \emph{Probability Alignment Score} (PAS) measures \emph{inferential} agreement while retaining uncertainty in the human and agent evidence: \revision{how closely that evidence agrees about an effect and its direction.} The \emph{Effect Consistency Score} (ECS) measures \emph{effect-magnitude} agreement: \revision{how closely the simulated effect sizes match the published human estimates.}

We test \textsc{HumanStudy-Bench} against 10 contemporary LLMs (open-weight and proprietary) under 4 standard agent designs (blank, role-play, demographic, contextualized backstory). \revision{The results reveal polarized test-level inferential agreement, limited effect-magnitude agreement, and configuration-dependent changes in agreement. Our evaluation asks three research questions: How closely do agents reproduce human behavioral effects? Where and how do simulated effects differ from human findings? How does agent design change agreement with human findings?}
\revision{Together, these analyses show not only how much agents align with human findings, but which effects align, how others differ, and how those patterns change across agent designs.}

\begin{table*}[t]
\centering
\color{pendingblue}
\caption{\textbf{Selected behavioral hypotheses and supporting evidence.} A subset of hypotheses from the 12 reconstructed studies. Effect sizes for the original and supporting studies follow the respective cited sources (Cohen's $d$ or $h$). See Appendix~\ref{app:study} for the full list.}
\label{tab:hypothesis-teaser}
\footnotesize
\setlength{\tabcolsep}{4pt}
\renewcommand{\arraystretch}{1.16}
\begin{tabularx}{\linewidth}{@{}>{\raggedright\arraybackslash}p{0.30\linewidth}*{4}{>{\raggedright\arraybackslash}X}@{}}
\toprule
\textbf{Main hypothesis} & \textbf{Original} & \textbf{Support 1} & \textbf{Support 2} & \textbf{Support 3} \\
\midrule
Gain framing favors the sure option; loss framing favors the risky option. &
\citet{Tversky1981TheFO}\newline $h=1.050$, $p<.001$ &
\citet{kuhberger1998influence}\newline $d=.31$ &
\citet{steiger2018meta}\newline $d=.52$ &
\citet{druckman2001using}\newline $h=.97$ \\
Ultimatum-game offers exceed dictator-game offers. &
\citet{Forsythe1994FairnessIS}\newline $d=1.576$, $p\approx0$ &
\citet{oosterbeek2004cultural}\newline $d\approx.40$ &
\citet{henrich2001search}\newline $d\approx.39$ &
\citet{engel2011dictator}\newline $h=1.12$ \\
Students estimate peers to be more comfortable with campus drinking practices than themselves. &
\citet{Prentice1993PluralisticIA}\newline $d=1.307$, $p<.0001$ &
\citet{schroeder1998exposing}\newline $d=.44$ &
\citet{suls2003pluralistic}\newline $d=1.04$ &
\citet{borsari2003descriptive}\newline $d=.70$ \\
\bottomrule
\end{tabularx}
\end{table*}

We make four contributions.
\begin{enumerate}[nosep,leftmargin=*]
    \item \revision{\textbf{A behavioral evaluation lens} that uses hypotheses screened against independent human evidence to make specific dimensions of human-likeness objectively testable, decomposable, and scalable.}
    \item \textbf{Two alignment metrics, PAS and ECS}, that quantify \emph{inferential} and \emph{effect-magnitude} agreement \revision{between simulated responses and published human findings, with PAS retaining uncertainty in the human and agent evidence.}
    \item \textbf{\textsc{HumanStudy-Bench}, an open platform} that reconstructs published human-subject experiments end-to-end and administers the test to user-configured agents, with an initial suite of 12 studies \revision{and support for community contributions.}
    \item \textbf{An empirical study} of 10 LLMs and 4 agent designs, generating 6,588 simulated participants for each agent configuration, \revision{identifying effect-specific discrepancies and the influence of agent design choices.}
\end{enumerate}

\section{Background \& Related Work}

\revision{\textbf{Behavioral evaluation with human experiments.}}
\revision{Prior work evaluates LLMs as simulated human populations through both observational and experimental comparisons. Survey- and dataset-based evaluations compare aggregate model responses with human responses across questions, tasks, and demographic groups, revealing misalignment, reduced response diversity, and distorted relationships between participant characteristics and responses~\citep{pmlr-v202-santurkar23a,tjuatja2024llmsexhibithumanlikeresponse,participants_nature,Bisbee_Clinton_Dorff_Kenkel_Larson_2024,hu2026simbench}. A complementary line of work administers human-subject experiments to simulated participants and examines whether population-level findings reappear. Turing Experiments establish this paradigm across classic experiments in behavioral economics, psycholinguistics, and social psychology~\citep{aher2023using}; subsequent studies evaluate cognitive effects, trust and economic behavior, prosocial biases, and larger collections of scenario-based and field experiments~\citep{binz2023using,xie2024trust,Kitadai_2025replaceeco,exploring_iclr2025,cui2025replication,chen2024field,ashokkumar2026predict}.}

\revision{Our unit of evaluation is the behavioral hypothesis, screened using subsequent independent human evidence. We distinguish direct support for the same or a closely matched contrast from related support for the broader behavioral effect, and score hypotheses with either form of support. For each scored hypothesis, we rerun the original statistical tests on agent responses and standardize agreement with the published human finding in inferential conclusion and, where comparable estimates are available, effect magnitude. This allows us to identify which specific hypotheses an agent reproduces, misses, reverses, or distorts in magnitude.}

\revision{\textbf{Role-playing and agent design.}}
\revision{LLMs simulate people by playing roles~\citep{shanahan2023roleplay}, and methods for specifying these roles fall into two groups. Prompt-based specifications describe a participant through demographics, backstories, personality profiles, or cognitive biases~\citep{argyle2023out,hu2024quantifyingpersonaeffectllm,lutz-etal-2025-prompt,moon2024virtualpersonaslanguagemodels,jiang2023evaluating,jiang-etal-2024-personallm,serapio2025psychometric,miotto-etal-2022-gpt,Cobra}. Data-based methods train or calibrate agents on character experiences, interviews, surveys, or behavioral records~\citep{shao2023character,park2024generativeagentsimulations1000,cao-etal-2025-specializing,suh2025languagemodelfinetuningscaled,kolluri2025finetuningllmshumanbehavior,xiaohongshu_socialworldmodel,binz2025foundation}. \textsc{HumanStudy-Bench} provides open, reusable, and extensible evaluation infrastructure for this line of research. Its reconstructed experimental protocols and evidence-screened behavioral hypotheses can be reused to evaluate new agent-specification methods, while additional studies and agent designs can be incorporated as the field develops.}

\Needspace{10\baselineskip}
\section{\textsc{HumanStudy-Bench}}
\revision{An \emph{agent configuration} pairs a base model with an \emph{agent design} (blank, role-play, demographic, or contextualized backstory); all configurations are evaluated under the same reconstructed protocols. This enables comparisons between agent designs within each reconstructed experimental setting.}

\newrevision{Working with domain experts, we curated 12 studies from approximately 100 candidates. Table~\ref{tab:hypothesis-teaser} shows illustrative behavioral hypotheses and independent evidence; Appendix~\ref{app:study} lists all reconstructions. Each agent configuration includes the same 6,588 simulated participants across the suite, with sample sizes set by the reconstructed studies; the original human samples range from tens to thousands of participants. These studies provide an initial set of behavioral dimensions for diagnostic evaluation. The platform supports adding studies and agent designs (Appendix~\ref{commnity}).}

\subsection{Reconstructing and Replaying Experimental Protocols}
\label{subsubsec:filter}
\revision{The human-in-the-loop \textbf{Filter--Extract--Execute} pipeline reconstructs each study, with outputs verified before deployment.}

\revision{\textbf{Filter.} We use follow-up evidence from independent human studies and meta-analyses to screen the behavioral hypotheses. We select studies with documented stimuli, instructions, conditions, quantifiable outcomes, and statistical analyses. Studies must be feasible to reconstruct without specialized physiological equipment. An LLM-assisted checklist supports screening; human reviewers assess the supporting evidence and make inclusion decisions. Every scored behavioral hypothesis has direct or related support from later independent human studies or meta-analyses; others are reconstructed but not scored (Appendix~\ref{app:finding-evidence}).}

\revision{\textbf{Extract.} We formalize the experimental design, hypotheses, statistical tests, and reported human outcomes, including statistics, $p$-values, and descriptive summaries. Participant sample sizes, demographics, and assignments are extracted where available. These define population-specific references and optional priors for agent specification.}

\newrevision{\textbf{Execute.} Study modules encode conditions, stimulus order, and task instructions. The engine presents trial instances to agents and stores their individual responses. Study-specific analyses then aggregate those responses into statistics comparable with the published human summaries; no original human individual records are required. Appendix~\ref{app:study} describes the reconstructions and adaptations, including text-based versions of laboratory tasks.}

\Needspace{10\baselineskip}
\subsection{\revision{Measuring Agreement on Behavioral Hypotheses}}
\label{subsubsec:evaluate}
\revision{Each study contains behavioral hypotheses, each operationalized by one or more statistical tests. We compare agent outputs with the corresponding published human findings using these analyses. PAS summarizes inferential agreement; ECS summarizes effect-magnitude agreement. The metrics provide complementary summaries of the scored effects: evidence strength depends on effect size, variability, and sample size.}

\paragraph{\revision{Probability Alignment Score (PAS).}}
\deletedrevision{A $p$-value describes how surprising a statistic is under a null in one sample, so comparing human and agent $p$-values does not measure whether they support the same finding. A direction-matching rate also treats weak evidence and strong evidence alike. PAS instead compares support for the positive, negative, and null states, retaining uncertainty about each result.}
\newrevision{We seek a continuous measure of whether agents reach the same inferential conclusions as the published human findings, while retaining uncertainty in each result. PAS compares the human and agent support for a hypothesis's positive, negative, and null states, then aggregates agreement across its tests.}
\newrevision{Formulating this as a Bayesian estimation problem:} \revision{in the binary case, a test-specific Bayes factor $\Lambda=BF_{10}$ is mapped to $\pi=P(H_1\mid D)=\Lambda/(1+\Lambda)$ under equal prior odds. Appendix~\ref{app:prior-choice} specifies the priors and approximations. Assuming the human and agent posterior states factorize, their agreement is}
\begingroup\color{black}
\begin{equation}
\mathrm{PAS}_j=\pi_{h,j}\pi_{a,j}+(1-\pi_{h,j})(1-\pi_{a,j}).
\end{equation}
\endgroup
\newrevision{$\mathrm{PAS}_j$ is the posterior probability that the human and agent hypothesis states agree, conditional on the observed data and the factorized posterior model above.}
\revision{For signed hypotheses we use the directional, three-state form, with posterior vectors over positive, negative, and null effects, $(H_+,H_-,H_0)$ (Appendix~\ref{app:generalization}):}
\begingroup\color{black}
\begin{equation}
\mathrm{PAS}_j=\pi_{h,+}\pi_{a,+}+\pi_{h,-}\pi_{a,-}+\pi_{h,0}\pi_{a,0}.
\label{eq:directional-pas}
\end{equation}
\endgroup
\revision{Thus, strong evidence for opposite directions receives low agreement even if both effects are significant.} \deletedrevision{Under this scoring rule, uniform human evidence over $K$ states gives agreement $1/K$ regardless of the agent posterior.}

\revision{To combine test-level PAS within a behavioral hypothesis, we take their arithmetic mean:}
\begingroup\color{black}
\begin{equation}
\mathrm{PAS}_{\mathrm{hypothesis}}=\frac{1}{M}\sum_{j=1}^{M}\mathrm{PAS}_j,
\end{equation}
\endgroup
\revision{Hypothesis scores are averaged within each study in the same way. Because each test-level PAS is the probability that the agent reaches the same conclusion as the human study on that test, a hypothesis or study score is the expected proportion of its tests on which agent and human conclusions agree. We then take the arithmetic mean of study scores for $\mathrm{PAS}_{\mathrm{global}}$, so studies with more hypotheses or larger samples do not dominate the benchmark (Appendix~\ref{app:aggregation_impl}). Scores are interpreted under the specified evidence model and study design.}

\paragraph{\revision{Effect Consistency Score (ECS).}}
\deletedrevision{Matching the direction of a finding leaves open whether agents reproduce its size. A raw effect-size MSE mixes outcome scales across studies; even on standardized effects, a single MSE value does not show whether disagreement reflects a changed pattern of effects or an overall shift or rescaling. ECS uses a bounded concordance score that exposes these two sources of disagreement.}
\newrevision{We also ask whether agents reproduce the magnitudes of human effects across tests. ECS compares standardized effect estimates and separates disagreement in their test-level pattern from overall differences in level and scale.}
\revision{We convert reported statistics to standardized effect estimates as described in Appendix~\ref{app:effect_size}, then apply Lin's concordance correlation coefficient~\citep{lawrence1989concordance}. Its unweighted form has a simple connection to mean squared error:}
\begingroup\color{black}
\begin{equation}
\mathrm{ECS}=\rho C_b=1-\frac{\mathrm{MSE}}{V}.
\end{equation}
\endgroup
\deletedrevision{Here $\rho$ measures correspondence in the pattern of effects, while $C_b$ detects overall level and scale mismatch; the signed mean difference indicates whether agents tend to over- or underestimate the human effects. Unlike Pearson correlation alone, concordance penalizes a uniformly rescaled effect vector.} \newrevision{MSE is the mean squared difference between standardized agent and human effects across tests. The normalizer $V$ is their expected squared difference if the two effect vectors were uncorrelated: the independence baseline. ECS therefore measures how much the observed squared difference falls below this baseline. The correlation $\rho$ tracks whether larger human effects also tend to be larger agent effects; $C_b$ penalizes shifts in the mean and differences in spread, which correlation alone misses. Appendix~\ref{app:aggregation_impl} gives the test-level weighted aggregate used in the benchmark.}
\revision{ECS measures agreement with the published numerical references. Appendix~\ref{app:effect_heterogeneity} examines sensitivity to test-family pooling and sampling noise under a specified model.}

\section{Experiment Setup}
\begin{table}\footnotesize
\centering
\caption{\textbf{Main leaderboard.} Probability Alignment Score (PAS) measures agreement about whether effects occur and in which direction; Effect Consistency Score (ECS) measures agreement in effect magnitude. Domain columns report PAS for individual cognition (Cog.), strategic interaction (Stra.), and social psychology (Soc.). Teal highlights the highest values and salmon the lowest. $^{\dagger}$Study~12 A4 uses Grok 4.3 (Grok 4.1 Fast was retired). Appendix~\ref{app:Moreexp} reports costs.}
\label{tab:pas-ecs-raw}
\setlength{\tabcolsep}{1.5pt}
\renewcommand{\arraystretch}{1.08}
\begin{tabular}{@{}l c c c c c c @{\hspace{5pt}\vrule\hspace{5pt}} l c c c c c c@{}}
\toprule
\textbf{Model} & \textbf{} & \textbf{Cog.} & \textbf{Stra.} & \textbf{Soc.} & \textbf{PAS} & \textbf{ECS} &
\textbf{Model} & \textbf{} & \textbf{Cog.} & \textbf{Stra.} & \textbf{Soc.} & \textbf{PAS} & \textbf{ECS} \\
\midrule
\multirow{4}{*}{\shortstack[l]{Claude\\Haiku\\4.5}}
 & A1 & $0.45$ & $0.25$ & $0.62$ & $0.440$ & $0.074$ &
\multirow{4}{*}{\shortstack[l]{GPT 5\\Nano}}
 & A1 & $0.43$ & \cellcolor{best3}{$0.62$} & $0.50$ & $0.515$ & \cellcolor{best3}{$0.259$} \\
 & A2 & \cellcolor{worst4}{$0.41$} & $0.26$ & $0.62$ & $0.432$ & \cellcolor{best4}{$0.228$} &
 & A2 & $0.55$ & $0.39$ & $0.56$ & $0.501$ & \cellcolor{best1}{$0.292$} \\
 & A3 & $0.47$ & $0.32$ & $0.49$ & $0.427$ & $0.206$ &
 & A3 & $0.53$ & \cellcolor{best1}{$0.65$} & \cellcolor{worst3}{$0.43$} & \cellcolor{best4}{$0.539$} & \cellcolor{best2}{$0.266$} \\
 & A4 & $0.43$ & \cellcolor{worst3}{$0.21$} & $0.55$ & \cellcolor{worst3}{$0.394$} & $0.127$ &
 & A4 & $0.57$ & \cellcolor{best2}{$0.63$} & $0.57$ & \cellcolor{best1}{$0.589$} & $0.206$ \\
\midrule
\multirow{4}{*}{\shortstack[l]{DeepSeek\\V3.2}}
 & A1 & $0.46$ & $0.21$ & \cellcolor{best4}{$0.63$} & $0.433$ & $0.104$ &
\multirow{4}{*}{\shortstack[l]{GPT\\OSS\\120b}}
 & A1 & \cellcolor{worst2}{$0.40$} & $0.22$ & $0.46$ & \cellcolor{worst1}{$0.359$} & \cellcolor{worst3}{$0.037$} \\
 & A2 & $0.48$ & $0.25$ & $0.55$ & $0.428$ & $0.148$ &
 & A2 & $0.44$ & $0.24$ & $0.51$ & \cellcolor{worst4}{$0.397$} & \cellcolor{worst4}{$0.039$} \\
 & A3 & $0.51$ & \cellcolor{worst4}{$0.21$} & \cellcolor{best3}{$0.63$} & $0.450$ & $0.115$ &
 & A3 & $0.50$ & $0.27$ & $0.49$ & $0.422$ & $0.123$ \\
 & A4 & $0.47$ & $0.22$ & \cellcolor{best2}{$0.65$} & $0.448$ & $0.134$ &
 & A4 & $0.49$ & $0.25$ & \cellcolor{best1}{$0.68$} & $0.472$ & $0.080$ \\
\midrule
\multirow{4}{*}{\shortstack[l]{Gemini 3\\Flash}}
 & A1 & $0.50$ & $0.51$ & $0.62$ & \cellcolor{best2}{$0.542$} & $0.096$ &
\multirow{4}{*}{\shortstack[l]{GPT\\OSS\\20b}}
 & A1 & $0.50$ & $0.34$ & $0.47$ & $0.437$ & $0.152$ \\
 & A2 & $0.51$ & $0.47$ & $0.44$ & $0.475$ & $0.058$ &
 & A2 & $0.47$ & $0.30$ & $0.45$ & $0.406$ & $0.137$ \\
 & A3 & \cellcolor{best2}{$0.62$} & $0.45$ & $0.55$ & \cellcolor{best3}{$0.539$} & $0.146$ &
 & A3 & $0.54$ & $0.42$ & \cellcolor{worst4}{$0.43$} & $0.464$ & $0.193$ \\
 & A4 & \cellcolor{best1}{$0.67$} & \cellcolor{worst2}{$0.20$} & $0.56$ & $0.476$ & $0.141$ &
 & A4 & $0.52$ & $0.33$ & $0.51$ & $0.453$ & $0.175$ \\
\midrule
\multirow{4}{*}{\shortstack[l]{Mistral\\Nemo}}
 & A1 & $0.56$ & $0.30$ & $0.52$ & $0.462$ & $0.114$ &
\multirow{4}{*}{\shortstack[l]{Qwen 3\\Next80b}}
 & A1 & $0.52$ & $0.49$ & $0.53$ & $0.514$ & \cellcolor{worst2}{$0.000$} \\
 & A2 & $0.55$ & $0.32$ & \cellcolor{worst1}{$0.41$} & $0.428$ & $0.095$ &
 & A2 & $0.50$ & $0.51$ & $0.53$ & $0.516$ & $0.159$ \\
 & A3 & $0.58$ & $0.27$ & $0.52$ & $0.458$ & $0.208$ &
 & A3 & $0.47$ & \cellcolor{best4}{$0.54$} & $0.52$ & $0.509$ & $0.112$ \\
 & A4 & \cellcolor{best3}{$0.61$} & $0.32$ & $0.53$ & $0.484$ & $0.205$ &
 & A4 & $0.55$ & $0.51$ & $0.50$ & $0.518$ & $0.096$ \\
\midrule
\multirow{4}{*}{\shortstack[l]{Gemma 4\\26b}}
 & A1 & \cellcolor{worst3}{$0.40$} & $0.49$ & $0.58$ & $0.489$ & $0.100$ &
\multirow{4}{*}{\shortstack[l]{Grok 4.1\\Fast}}
 & A1 & $0.44$ & $0.26$ & $0.51$ & $0.404$ & $0.080$ \\
 & A2 & \cellcolor{worst1}{$0.37$} & $0.49$ & $0.56$ & $0.473$ & $0.092$ &
 & A2 & $0.45$ & $0.24$ & \cellcolor{worst2}{$0.43$} & \cellcolor{worst2}{$0.372$} & $0.050$ \\
 & A3 & $0.46$ & $0.50$ & $0.57$ & $0.511$ & $0.146$ &
 & A3 & $0.55$ & $0.23$ & $0.57$ & $0.449$ & \cellcolor{worst1}{$-0.008$} \\
 & A4 & $0.47$ & $0.24$ & $0.61$ & $0.442$ & $0.130$ &
 & A4 & \cellcolor{best4}{$0.59$} & \cellcolor{worst1}{$0.19^{\dagger}$} & $0.48$ & $0.421^{\dagger}$ & $0.182^{\dagger}$ \\
\bottomrule
\end{tabular}
\end{table}

\subsection{\revision{Shared Benchmark Evaluation}}
\textbf{Human Studies.}
We evaluate against 12 human-subject studies (Appendix~\ref{app:study}) spanning individual cognition, strategic interaction, and social psychology, scoring the 45 behavioral hypotheses with direct or related follow-up support (Appendix~\ref{app:finding-evidence}). The individual cognition studies focus on cognitive biases and heuristic judgment~\citep{Ross1977TheC,Jacowitz1995MeasuresOA,Tversky1981TheFO,Kahneman1972SubjectivePA}. Strategic interaction studies are drawn from paradigms in game theory~\citep{nagel1995unraveling,Shafir1992ThinkingTU,Forsythe1994FairnessIS,Berg1995TrustRA}. Social psychology studies examine social cognition, social norms, and group behavior~\citep{article,asch1946forming,gagnon1996discrimination,Prentice1993PluralisticIA}.

\textbf{Models and Inference Settings.}
\newrevision{We evaluate 10 contemporary models, including open-weight (e.g., Mistral, DeepSeek, Qwen) and proprietary APIs (e.g., Claude, GPT, Gemini, Grok). Model details are in Appendix~\ref{app:model-id}. All models use a temperature $T=1.0$ to induce the behavioral variance needed for population simulation. Response-bootstrap standard errors are reported in Appendix~\ref{app:se}; these quantify resampling uncertainty within the generated agent responses.}

\textbf{Agent Design Variants.}
These variants are exploratory probes of how common forms of participant specification affect agreement with human findings. We use them to characterize sensitivity to agent design rather than to identify an optimal simulation method.
(1) \textit{Blank (A1)} uses the base model with no additional specification.
(2) \textit{Role-Play (A2)} instructs the model to act as a human participant in a psychological study, without specific attributes.
\revision{(3) \textit{Demographic (A3)} adds attributes (e.g., age, gender, occupation) to the role-playing instruction, based on reported participant demographics where available.}
(4) \textit{Contextualized Backstory (A4)} augments demographics with a rich natural language narrative about the agent's life history, personality, and daily context.
See Appendix~\ref{app:version} for details.

\textbf{Evaluation Protocol.}
\newrevision{The reconstructed protocols assign simulated participants to the original experimental conditions and collect their responses. By default, each agent configuration uses the same participant counts and condition allocation as the original human study. Study-specific evaluators parse these responses, recompute the relevant sample statistics, and compare them with published human summaries. We report the Probability Alignment Score (PAS; $[0,1]$) for inferential agreement and the Effect Consistency Score (ECS; $[-1,1]$) for effect-magnitude agreement ($1$ indicates exact agreement on the scored summaries).}

\subsection{\revision{Analyses for the Three Research Questions}}
\label{sec:validation-setup}
\revision{RQ1 uses PAS and ECS to measure how closely agents reproduce the direction and magnitude of the selected human behavioral effects. For RQ2, the unit of comparison is a statistical test evaluated under one agent configuration. We first ask whether the agent responses support the same effect direction as the published human result and then, for direction-matched comparisons with comparable effect estimates, whether the simulated effect is substantially weaker or stronger. For RQ3, our primary analysis pairs the same model and behavioral hypothesis across adjacent agent designs: A1 to A2, A2 to A3, and A3 to A4. We balance these paired changes across studies and resample studies for exploratory uncertainty intervals.}

\section{Results}
\label{sec:results}
\revision{This section organizes the results around the three research questions.}
\begin{figure}
\centering
\subfigure[Effect-size correspondence.]{\includegraphics[width=0.47\linewidth]{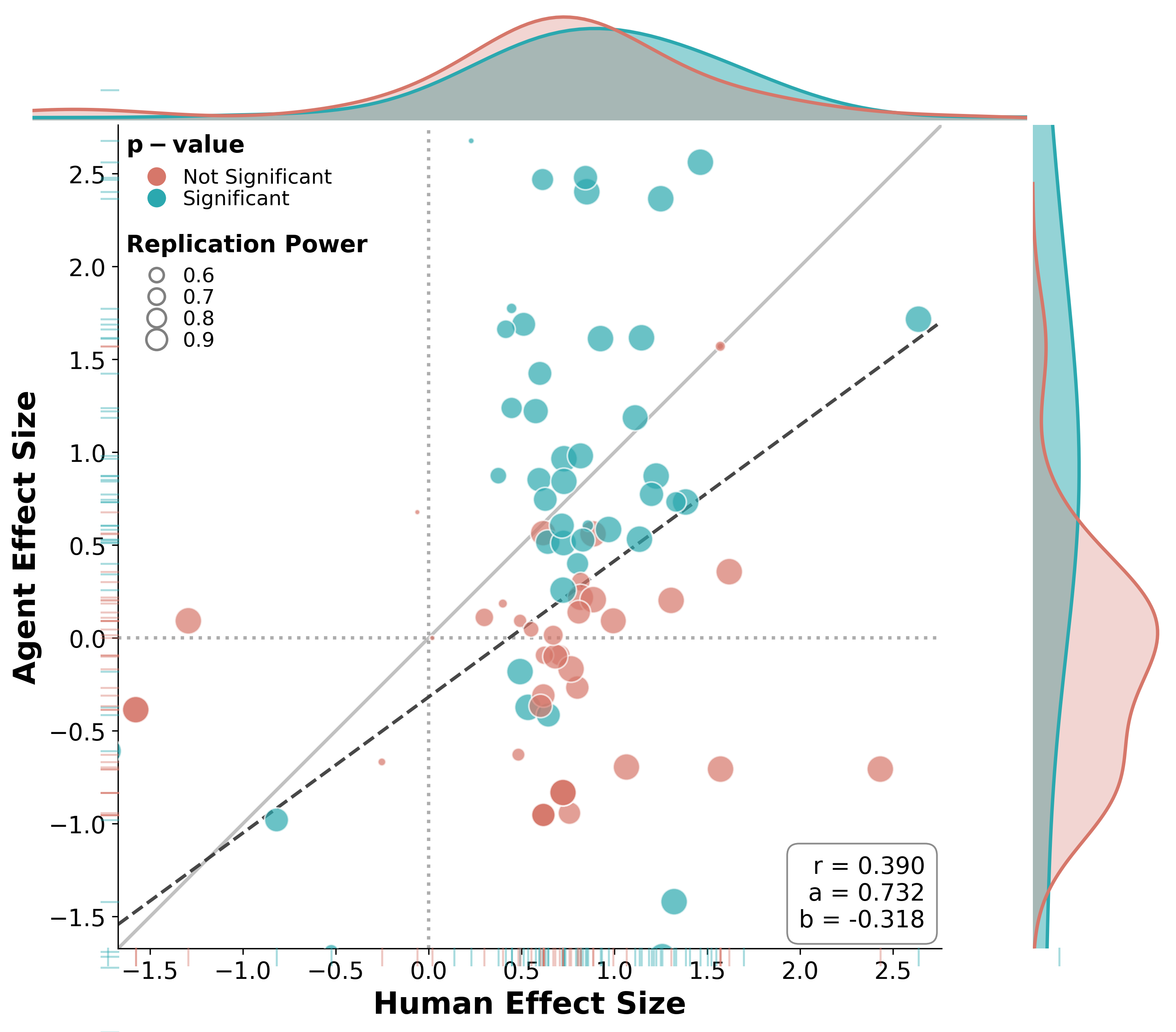}\label{fig:distributional-analysis}}
\hfill
\subfigure[Test-level PAS.]{\includegraphics[width=0.47\linewidth]{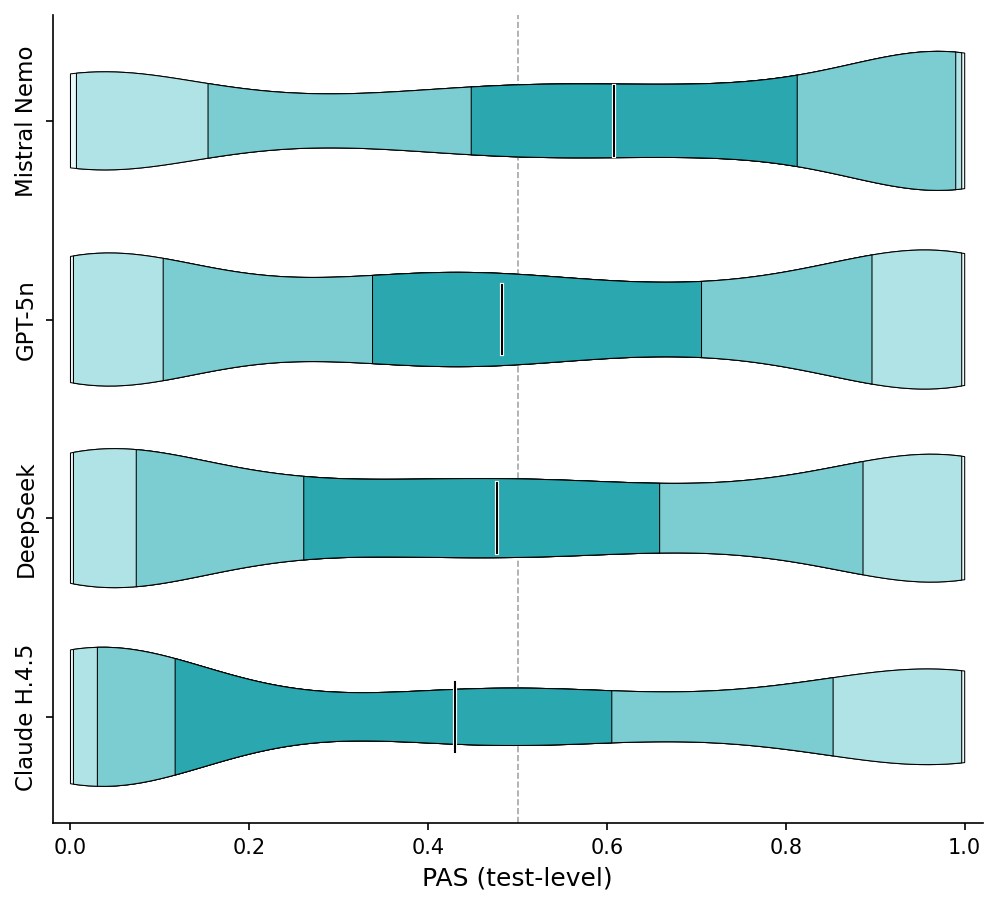}\label{fig:pas-violin}}
\caption{\revision{\textbf{Complementary views of agreement.} (a) Effect-size correspondence for GPT-5 Nano A2. The diagonal marks exact agreement; the dashed line is a fitted regression. (b) Test-level PAS distributions. Many agent scores lie near the extremes of inferential agreement on the scored hypotheses.}}
\label{fig:per-test-alignment}
\end{figure}
\subsection{\revision{RQ1: How Closely Do Agents Reproduce Human Behavioral Effects?}}
\label{sec:rq1}
Overall, agents sometimes reproduce whether a human effect occurs and in which direction, but reproduce its magnitude much less reliably.\\
\textbf{\revision{Inferential agreement.}}
\revision{Across individual tests, agreement is often polarized: 25\% of test-level scores exceed $0.9$ and 31\% fall below $0.1$ (Figure~\ref{fig:pas-violin}). Agreement also varies across domains and agent configurations. For example, Claude Haiku 4.5 with role-play instructions achieves a PAS of $0.62$ on the social-psychology tasks but only $0.26$ on the strategic-interaction tasks (Table~\ref{tab:pas-ecs-raw}). The leading model--design configuration differs by domain: Gemini 3 Flash A4 scores highest in cognition, GPT-5 Nano A3 in strategic interaction, and GPT OSS 120b A4 in social psychology (Table~\ref{tab:pas-ecs-raw}). Reproducing one set of human findings therefore does not imply reliable agreement on others.}\\
\textbf{\revision{Effect-magnitude agreement.}}
\revision{Agents reproduce effect magnitudes less reliably than effect directions. Even the strongest configuration reaches an ECS of only $0.292$ (GPT-5 Nano with role-play instructions), indicating limited correspondence between simulated and human effect magnitudes (Table~\ref{tab:pas-ecs-raw}). Figure~\ref{fig:distributional-analysis} illustrates why: some simulated effects lie close to the line of exact agreement, but many fall well above or below it. For GPT-5 Nano with role-play instructions, the simulation understates some published effects and overstates others rather than shifting all effects in the same direction. Thus, an agent may correctly reproduce whether an effect is positive or negative while still giving a misleading account of how large it is.}

\begin{figure}
\centering
\includegraphics[width=\linewidth]{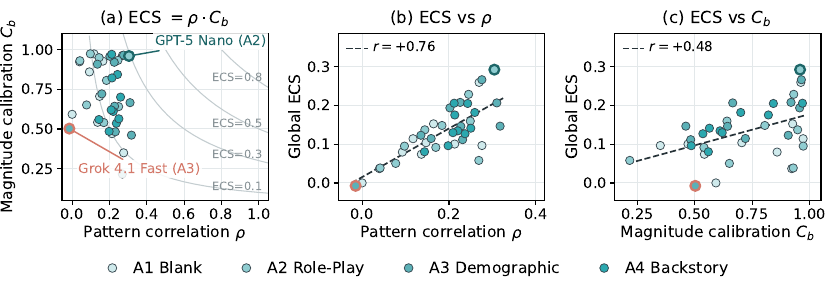}
\caption{\revision{\textbf{ECS decomposition.} (a) Pattern correlation $\rho$ and calibration factor $C_b$ for evaluated agents. (b,c) Their associations with ECS. Low $\rho$ indicates limited correspondence in the relative magnitudes of the scored effects.}}
\label{fig:ecs-decomposition}
\end{figure}

\revision{The same weak correspondence remains when effect magnitudes are compared separately within different statistical test families. Figure~\ref{fig:ecs-decomposition} summarizes whether ECS is limited by disagreement in the relative pattern of effects or by differences in their overall scale. A mixture of the ten models' responses agrees with the human findings at least as well as most single models (Appendix~\ref{app:hypothesis_details}); sensitivity analyses are in Appendix~\ref{app:effect_heterogeneity}.}

\subsection{\newrevision{RQ2: Where and How Do Simulated Effects Differ from Human Findings?}}
\label{sec:rq2}
\newrevision{We observe two common forms of disagreement between simulated and human results. First, agents often fail to reproduce the direction of the human effect. Among comparisons where the human study reports a positive or negative effect, 50\% of the simulated results show no effect or an effect in the opposite direction. Second, even when the simulated effect points in the same direction as the human effect, its magnitude is often substantially different. After excluding human effects too close to zero for a meaningful comparison, 54\% of simulated effects are less than half or more than twice the published estimate, with overestimation occurring more often than underestimation. Table~\ref{tab:effect-discrepancies} summarizes these two forms of disagreement.}
\begin{table}[H]
\footnotesize
\centering
\color{pendingblue}
\caption{\textbf{Patterns of behavioral discrepancy.} A simulated effect can differ from a published human finding in its direction or in its magnitude. Appendix~\ref{app:effect-cases} reports the response-level behaviors.}
\label{tab:effect-discrepancies}
\setlength{\tabcolsep}{5pt}
\renewcommand{\arraystretch}{1.18}
\begin{tabularx}{\linewidth}{@{}>{\raggedright\arraybackslash}p{0.22\linewidth}>{\raggedright\arraybackslash}X@{}}
\toprule
\textbf{Pattern} & \textbf{Observable signature in simulated responses} \\
\midrule
Direction mismatch & The agent responses provide no evidence for the published effect, or support the opposite direction. \\
\addlinespace[2pt]
Magnitude mismatch & The direction is retained, but the simulated effect is substantially larger or smaller than the published estimate. \\
\bottomrule
\end{tabularx}
\end{table}

\newrevision{For example, the framing task illustrates a direction mismatch: human choices reversed between gain and loss wording, while agent choices often favored the certain option under both. The peer-comfort task illustrates attenuation: the median simulated self--peer gap was 0.63 points, compared with 1.71 points in the human study. Appendix~\ref{app:effect-cases} provides the classification rules and response-level examples.}

\subsection{\revision{RQ3: How Does Agent Design Change Agreement with Human Findings?}}
\label{sec:rq3}
\revision{We observe two patterns in how agent configurations change agreement with human findings. First, simply instructing a model to act as a human participant does not reliably improve agreement with the human findings. Moving from no participant specification (A1) to role-playing instructions (A2) slightly lowers PAS on average. Across the 45 behavioral hypotheses, PAS increases by more than $0.02$ for 15 hypotheses but decreases by more than $0.02$ for 15, with no clear overall improvement.}

\begin{figure}
\centering
\includegraphics[width=\linewidth]{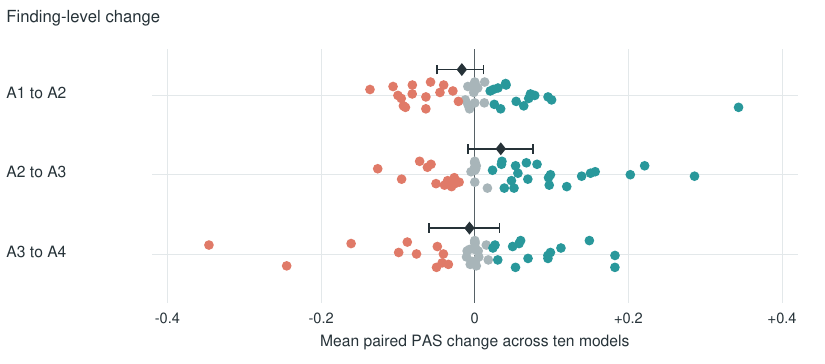}
\caption{\textbf{Hypothesis-level changes across agent designs.} Each dot shows, for one behavioral hypothesis, the average change in PAS across the ten models when one layer of participant description is added. Each comparison covers 45 hypotheses from twelve studies. Black diamonds show averages that give each study equal weight, and the accompanying 95\% intervals are obtained by resampling studies. Teal marks hypotheses whose PAS increases by more than $0.02$, salmon marks decreases of more than $0.02$, and gray marks smaller changes.}
\label{fig:rq3-finding-changes}
\end{figure}

\revision{Second, adding more participant information also produces mixed results. Adding demographics to the role-playing instructions (A2 to A3) raises PAS by $0.034$ on average, with A3 exceeding A2 for eight of ten models (Wilcoxon $p=0.010$; Appendix~\ref{app:hypothesis_details}). Adding contextualized backstories (A3 to A4) produces approximately no average change. This near-zero average masks changes in opposite directions: PAS increases by more than $0.02$ for 15 hypotheses and decreases by more than $0.02$ for 11 (Figure~\ref{fig:rq3-finding-changes}). Thus, richer participant descriptions improve agreement for some human findings but reduce it for others.}

\paragraph{Implications for participant simulation.}
Our results suggest that agreement between simulated and human populations depends on both the agent design and the behavioral hypothesis being tested: a design that reproduces one effect may miss another, and changing the design can improve agreement on some hypotheses while worsening it on others. \textsc{HumanStudy-Bench} supports these comparisons through reusable experimental protocols and makes it easier to add new hypotheses by reconstructing studies from papers and experimental materials. This open infrastructure enables the community to build on accumulated behavioral knowledge and repeatedly assess simulated populations as models and agent designs evolve.

\section{Conclusion}
\label{sec:conclusion}
\revision{Behavioral hypotheses provide a concrete unit for testing whether simulated participants reproduce documented human effects. \textsc{HumanStudy-Bench} operationalizes this idea by reconstructing experimental protocols from human studies, administering them to LLM agents serving as silicon participants, and comparing the resulting population-level effects with the corresponding published human findings, hypothesis by hypothesis. This hypothesis-level view makes the behavioral assumptions behind AI-based social simulations explicit, testable, and traceable to particular effects. Each scored test remains linked to the agent responses that produced it, allowing disagreement to be inspected at the level of choices and judgments. Through \textsc{HumanStudy-Bench}, we hope to turn social science experiments and their behavioral hypotheses into open, reusable resources for assessing silicon participants in social simulations.}

\bibliographystyle{iclr2027_conference}
\bibliography{references}
\clearpage
\appendix
\clearpage
\section*{AI Use Statement}
Generative AI enters this work in two ways: as a component of the \textsc{HumanStudy-Bench} pipeline, and as an aid in conducting the research and writing the paper. The LLMs evaluated as simulated participants are the objects of study.

\paragraph{AI in the \textsc{HumanStudy-Bench} pipeline.} The pipeline has two stages. \emph{Study construction (Filter--Extract--Execute).} An LLM from the Gemini 3 Flash family screens candidate papers against the inclusion checklist, extracts experimental designs, hypotheses, and reported statistics, and drafts the configuration and response-parsing code for each study. Human reviewers make all inclusion decisions, check every extracted parameter against the source paper, and cross-review and test the generated code before deployment. \emph{Simulation and evaluation.} Agents answer the reconstructed trials. For the A4 design, participant backstories are generated by Gemini 3 independently of the experimental task, and reviewers inspect them for task-specific answers. A DeepSeek formatter model rewrites free-text agent answers into the structured answer format; it is instructed to extract only values present in the response and not to supply missing answers. PAS and ECS are then computed by fixed statistical code; no LLM judges agreement between agent and human results.

\paragraph{AI in conducting the research and writing the paper.} We used generative AI to write plotting code, edit figures, search related works, assist the targeted literature check that classifies follow-up evidence for each behavioral hypothesis (Appendix~\ref{app:finding-evidence}), edit the manuscript for readability, provide methodological feedback, and check mathematical claims that the authors formulated and verified. We did not use AI to propose any research ideas or hypotheses, write proofs, or conduct qualitative or thematic analysis. The authors made the evidence classifications, checked citations against source papers, tested AI-assisted code, and reviewed all AI-assisted text and figures. We take responsibility for the final paper and artifacts, including content produced with the aid of AI.

\section{Metrics Theoretical Foundation}
\label{app:metrics}
\subsection{Metric Intuition: What Are We Measuring?}
\label{app:intuition}
PAS summarizes agreement between posterior states of specified hypotheses; ECS summarizes concordance between selected effect estimates. The metrics are related but answer different questions. A high PAS need not imply accurate magnitudes, and matching selected effect estimates does not establish the fidelity of the complete response distribution.

\subsection{Posterior Factorization Assumption}
\label{app:conditional}
For human and agent hypothesis states $\theta_h,\theta_a$, the probability-of-agreement interpretation assumes
\begin{equation}
P(\theta_h,\theta_a\mid D_h,D_a)=P(\theta_h\mid D_h)P(\theta_a\mid D_a).
\end{equation}
Separate sampling and independent priors with factorized likelihoods provide one model under which this holds.

\subsection{Foundation of Probability Alignment Score}
\label{app:theory}
Let $\mathrm{PAS}^*=\mathbf{1}(\theta_h=\theta_a)$. Under squared-error loss, its posterior mean minimizes posterior expected loss. With the factorization above and $K$ mutually exclusive hypothesis states,
\begin{equation}
\widehat{\mathrm{PAS}}=\mathbb{E}[\mathrm{PAS}^*\mid D_h,D_a]=\sum_{k=1}^K\pi_{h,k}\pi_{a,k}.
\end{equation}
If the human posterior is uniform over $K$ states, it gives no preferred reference conclusion and the agreement score equals $1/K$ for any agent posterior. Such a test cannot distinguish agents on inferential agreement.

For two states, this becomes $\pi_h\pi_a+(1-\pi_h)(1-\pi_a)$; the implemented directional extension is described in Appendix~\ref{app:generalization}.

For the binary evidence transform, equal prior odds imply $\pi=\Lambda/(1+\Lambda)$, where $\Lambda=BF_{10}$. Equal odds are a modeling choice; effect-size priors and approximations still affect the resulting probabilities. Appendix~\ref{app:prior-choice} reports a sensitivity analysis for the tested JZS scales. The continuous transformation avoids a discontinuous significance threshold, and the data-based directional split (Appendix~\ref{app:generalization}) keeps the three-state PAS continuous when an estimate crosses zero.

\section{Methodological Rationale and Aggregation}
\label{app:methodology}

\subsection{Methodological Rationale}
\label{app:rationale}
Our aggregation strategy and metric formulation diverge from standard meta-analytic approaches. We explicitly contrast our choices with alternative methodologies below.

\paragraph{Benchmarking vs. Meta-Analysis.}
While our framework aggregates results across multiple studies, we intentionally employ unweighted averaging rather than the precision-weighted averaging (inverse-variance weighting) typical of meta-analysis. This decision is grounded in two primary distinctions:

\begin{itemize}
    \item \textbf{Task Independence vs. Parameter Estimation:} Meta-analysis assumes that different studies estimate a shared biological parameter (e.g., a ``true'' population effect size). In contrast, our goal is benchmarking: evaluating an agent's general capability across a diverse suite of distinct tasks. Weighting by inverse variance would allow a single study with high statistical power to dominate the aggregate score, obscuring the agent's failure on smaller but equally critical tasks.

    \item \textbf{Avoiding Simulation Artifacts:} Although agent sample size ($N_{agent}$) can be controlled in a simulation, it is fixed by each reconstructed protocol in our evaluation. Precision weighting would allow larger simulated samples to dominate the benchmark score. Equal study weights prevent simulation volume from determining aggregation weights; PAS still uses the fixed sample size of each test when assessing its evidence.
\end{itemize}

\subsection{Metric Selection Rationale}
\label{app:metric_selection}
PAS focuses on evidence for specified hypotheses, while effect-size and distributional metrics answer complementary fidelity questions. Response variance can itself be behaviorally meaningful. Reported statistical summaries permit evaluation where original human microdata are unavailable, but limit checks of individual-level distributions. A common score range supports summarization across tests without removing differences in their statistical models or power.

\subsection{Aggregation Implementation}
\label{app:aggregation_impl}
Scores are aggregated across the hierarchy (Test $\to$ Behavioral Hypothesis $\to$ Study $\to$ Benchmark) using the following scoring convention:

\begin{enumerate}
    \item \textbf{Hypothesis \& Study Level (Arithmetic Mean):} We average test-level PAS within each behavioral hypothesis and hypothesis-level PAS within each study:
    \begin{equation}
        \mathrm{PAS}_{\mathrm{hypothesis}}=\frac{1}{M}\sum_{j=1}^{M}\mathrm{PAS}_j.
    \end{equation}
    Because each test-level PAS is the probability that the agent reaches the same conclusion as the human study on that test, each mean is the expected proportion of tests on which agent and human conclusions agree. Unlike averages on the log-odds scale, they are not dominated by scores near zero or one and need no clipping constant.

    \item \textbf{Benchmark Level (Arithmetic Mean):} We compute the unweighted arithmetic mean of study-level PAS. Given the heterogeneity of the studies---which span diverse cognitive and social domains---we treat each study as an independent unit of capability. This approach ensures equal representation across domains and prevents any single study with distinct statistical properties (e.g., large $N$) from dominating the global benchmark score.
\end{enumerate}
For ECS, only tests with recoverable agent and human effect pairs enter. Each such test $i$ in hypothesis $q$ of study $s$ receives weight $\tilde w_i=1/(K H_s T_{sq})$, where $K$ is the number of studies with at least one such test, $H_s$ is the number of hypotheses in study $s$ with such tests, and $T_{sq}$ is the number of such tests in hypothesis $q$. The weights sum to one, so each study contributes $1/K$ and each hypothesis within a study an equal share, matching the aggregation of PAS. With weighted means $\bar\delta$ and centered effects $u_{a,i},u_{h,i}$, the implemented concordance is
\begin{equation}
\mathrm{ECS}_{\mathrm{global}}=\frac{2\sum_i\tilde w_i u_{a,i}u_{h,i}}{\sum_i\tilde w_i u_{a,i}^2+\sum_i\tilde w_i u_{h,i}^2+(\bar\delta_a-\bar\delta_h)^2}.
\end{equation}
For the unweighted form in the main text, $\mathrm{MSE}=E[(a-h)^2]$. Expanding it gives $\sigma_a^2+\sigma_h^2-2\sigma_{ah}+(\mu_a-\mu_h)^2$. Thus $V=\sigma_a^2+\sigma_h^2+(\mu_a-\mu_h)^2$ is the expected squared difference when the covariance is zero, and $\mathrm{ECS}=1-\mathrm{MSE}/V$.

\section{Metrics Implementation Details}
\label{app:implementation}

\subsection{Evidence Transformation (Priors)}
\label{app:prior-choice}
To compute posterior probabilities $\pi$, we calculate Bayes Factors ($BF_{10}$) using priors tailored to the test type. For t-tests and ANOVA, we employ the JZS prior (Cauchy distribution on effect size) with default scales $r=\sqrt{2}/2$ and $r=0.5$, respectively \citep{rouder2009bayesian,rouder2012default}. These account for the majority of the tests. For contingency tables, we utilize a BIC-style approximation ($BF_{10} \approx \exp((\chi^2 - \text{df}\ln n)/2)$), while binomial tests use an exact conjugate Beta-Binomial prior (Beta(1,1)).

\paragraph{Sensitivity Analysis.}
\label{app:sensitivity}
To ensure that the benchmark rankings are driven by agent capability rather than specific prior choices in the evidence transformation, we conducted a sensitivity analysis on the prior scale $r$ used in the JZS Bayes Factor computation. The default value $r=0.707$ assumes a medium effect size distribution. We re-evaluated the raw responses of the forty agent configurations on the scored behavioral hypotheses, varying $r$ from $0.5$ (small effects) to $1.0$ (large effects).

As shown in Table \ref{tab:sensitivity}, while the absolute magnitude of the posterior probabilities shifts slightly with the prior width, the relative ranking of agents remains highly stable (Spearman's $\rho \geq 0.998$; mean PAS changes by at most $0.006$).

We also varied the structural choices of PAS, for human and agent posteriors alike (lower block of Table~\ref{tab:sensitivity}): prior odds of $1/3$ and $3$; the sign probability $P(\operatorname{sign}\delta\mid D,H_1)$ taken under the JZS prior of the Bayes factor rather than a noninformative prior (t tests and single-df ANOVAs); a sign probability of $1$ for the observed sign; and Gunel--Dickey Bayes factors with uniform priors \citep{gunel1974bayes} for contingency tables whose counts are available to the evaluator (64\% of $\chi^2$ tests). Rankings remain stable (Spearman's $\rho \geq 0.95$). A3 exceeds A2 and social exceeds strategic PAS under every variant.

\begin{table}[h]
\centering
\caption{\textbf{Sensitivity of agent rankings to PAS choices.} Upper block: Cauchy prior scale $r$. Lower block: one structural choice changed at a time. $\rho$ is the Spearman correlation of the forty configuration PAS values with the default.}
\label{tab:sensitivity}
\small
\begin{tabular}{lcccc}
\toprule
\textbf{Setting} & \textbf{Spearman's $\rho$} & \textbf{Mean $\Delta$ PAS} & \textbf{Max $\Delta$ PAS} & \textbf{Status} \\
\midrule
$r=0.500$ & 0.9991 & 0.002697 & 0.005821 & Stable \\
$r=0.600$ & 0.9994 & 0.001347 & 0.002826 & Stable \\
$r=\mathbf{0.707}$ (Default) & \textbf{1.0000} & \textbf{0.0000} & \textbf{0.0000} & \textbf{Baseline} \\
$r=0.800$ & 0.9992 & 0.000967 & 0.002217 & Stable \\
$r=0.900$ & 0.9991 & 0.002053 & 0.004436 & Stable \\
$r=1.000$ & 0.9989 & 0.003069 & 0.006488 & Stable \\
\midrule
Prior odds $1/3$ & 0.9629 & 0.016529 & 0.039455 & Stable \\
Prior odds $3$ & 0.9537 & 0.032426 & 0.058871 & Stable \\
JZS sign probability & 0.9998 & 0.000401 & 0.001323 & Stable \\
Sign probability $=1$ & 0.9878 & 0.004973 & 0.019651 & Stable \\
Gunel--Dickey & 0.9976 & 0.007902 & 0.012221 & Stable \\
\bottomrule
\end{tabular}
\end{table}

\subsection{Generalization to Multiple Hypotheses}
\label{app:generalization}
Theoretically, PAS generalizes to $K$ hypotheses via the inner product of posterior vectors $\vec{\pi}_h \cdot \vec{\pi}_a$. In our implementation, we specifically operationalize this as a 3-way split ($H_+, H_-, H_0$). The resulting score is the dot product of the agent and human posterior vectors over these three outcome categories:
\begin{equation}
    S = \pi_{h+}\pi_{a+} + \pi_{h-}\pi_{a-} + \pi_{h0}\pi_{a0}
\end{equation}
Each test compares $H_0:\delta=0$ with $H_+:\delta>0$ and $H_-:\delta<0$ (for binomial tests, $\delta=p-p_0$). With equal prior odds for $H_0$ and $H_1=H_+\cup H_-$, and since $\pi_\pm=P(H_1\mid D)\,P(\operatorname{sign}\delta=\pm\mid D,H_1)$,
\begin{equation}
    \pi_0=\frac{1}{1+BF_{10}},\qquad \pi_\pm=\frac{BF_{10}}{1+BF_{10}}\,P(\operatorname{sign}\delta=\pm\mid D,H_1).
\end{equation}
For a prior symmetric about the null value this matches the one-sided Bayes factors, $BF_{\pm0}=2\,BF_{10}\,P(\operatorname{sign}\delta=\pm\mid D,H_1)$ \citep{morey2014simple}. By default, $P(\operatorname{sign}\delta\mid D,H_1)$ is the sign posterior under a noninformative prior: the Student-$t$ posterior for $t$ tests, single-df $F$ tests, and correlations (via their $t$ statistic), a normal approximation with $z=\sqrt{\chi^2}$ for $2\times2$ tables and $z$-scored statistics, and the Beta$(1,1)$ posterior mass on the observed side of $p_0$ for binomial tests. Taking it under the JZS prior of the Bayes factor instead changes PAS by at most $0.0013$ (Table~\ref{tab:sensitivity}).
Human and agent posteriors of a test use the same test and prior, so each test-level PAS is the probability that both fall in the same state. When the hypothesis has no sign, as for tests with more than one numerator degree of freedom, PAS compares only the effect and no-effect states, $\pi_{h1}\pi_{a1}+\pi_{h0}\pi_{a0}$ with $\pi_1=1-\pi_0$.

\subsection{Standardized Effect Size Recovery}
\label{app:effect_size}
We recover Cohen's $d$ from reported statistics using the following conversions:
\begin{itemize}
    \item \textbf{T-family:} Independent t-tests use $d = t\sqrt{(n_1+n_2)/(n_1 n_2)}$; paired/one-sample tests use $d = t/\sqrt{n}$. F-tests ($df_1=1$) are converted to $t$-equivalents ($t=\sqrt{F}$) and processed similarly.
    \item \textbf{Correlation-family:} Pearson's $r$, Fisher's $z$, and Mann-Whitney $U$ (via rank-biserial $r_{rb}$) are converted to $d$ using the relationship $d = 2r/\sqrt{1-r^2}$.
    \item \textbf{Discrete:} $2 \times 2$ contingency tables are converted via the Log Odds Ratio ($d \approx \ln(OR)\sqrt{3}/\pi$). Binomial proportions use Cohen's $h = 2\arcsin\sqrt{p} - 2\arcsin\sqrt{p_0}$ against the null proportion $p_0$.
\end{itemize}

\begingroup\color{pendingblue}
\section{Agent Design Variants}
\label{app:version}

All four designs receive the same study trial; only the agent specification changes. The role and demographic fields are assigned before the trial response is generated.

\subsection{Blank (A1)}
\label{subsec:v1}
No additional system-level participant description is provided.

\subsection{Role-Play (A2)}
\label{subsec:v2}
The system message asks the model to participate in a psychology experiment as a human participant, without demographic attributes.

\subsection{Demographic (A3)}
\label{subsec:v3}
The role instruction is supplemented with available age, gender, and education or background attributes sampled from the study's reported participant profile. The model is asked to follow the trial instructions and answer in the required format.

\subsection{Contextualized Backstory (A4)}
\label{subsec:v4}
The demographic profile is extended with a generated narrative about everyday life, relationships, routines, and personality. The narrative is placed in the system message before the same trial instructions. Backstory generation is independent of the specific experimental problem, so the specification adds participant context rather than a suggested answer.
\endgroup

\begingroup\color{pendingblue}
\section{Study Reconstructions}
\label{app:study}

The original papers supply protocols and reported human statistics, not participant-level human records. Each study module generates agent trials, stores individual agent responses, and computes study-specific summaries from those responses. The summaries are compared with the published human references. Table~\ref{tab:appendix_studies} describes the task in each reconstruction and the contrast it tests; adaptations to text-based interaction remain part of the interpretation of every comparison.

\begin{table}[!htbp]
\caption{\textbf{What each reconstructed study asks a simulated participant to do.} Studies are grouped by domain. The last column names the manipulation and the measured outcome behind the scored behavioral hypotheses (Table~\ref{tab:finding-evidence}). Human references are published summary results; original human microdata are not used.}
\label{tab:appendix_studies}
\centering\footnotesize\setlength{\parskip}{0pt}
\setlength{\tabcolsep}{4pt}
\renewcommand{\arraystretch}{1.0}
\begin{tabular}{@{}>{\raggedright\arraybackslash}p{0.2\linewidth}>{\raggedright\arraybackslash}p{0.5\linewidth}>{\raggedright\arraybackslash}p{0.235\linewidth}@{}}
\toprule
\textbf{Study} & \textbf{Task for the simulated participant} & \textbf{Manipulation $\rightarrow$ outcome} \\
\midrule
\multicolumn{3}{@{}l}{\textit{Individual cognition}} \\\addlinespace[3pt]
\textbf{False consensus} (001)\par \citet{Ross1977TheC} & Reads short dilemmas (e.g., whether to walk around campus wearing an advertising sign), chooses an action, estimates how many peers would choose each option, and rates the traits of people who would. & Own choice $\rightarrow$ estimated peer agreement and trait ratings \\\addlinespace[3pt]
\textbf{Anchoring} (002)\par \citet{Jacowitz1995MeasuresOA} & Says whether an uncertain quantity is above or below a given number, then estimates it and rates confidence. Variants call the anchor uninformative or ask for willingness to pay. & High vs.\ low anchor $\rightarrow$ numeric estimates \\\addlinespace[3pt]
\textbf{Framing} (003)\par \citet{Tversky1981TheFO} & Chooses between options in classic decision problems: the Asian disease problem, concurrent gambles, a pseudo-certain prize, and a lost theater ticket vs.\ lost cash. & Gain vs.\ loss wording and related reframings $\rightarrow$ choices \\\addlinespace[3pt]
\textbf{Representativeness} (004)\par \citet{Kahneman1972SubjectivePA} & Judges how likely samples are: birth-order sequences, school-program majorities, marble distributions, averages from small and large samples, and heights. & How representative a sample looks, or its size $\rightarrow$ probability judgments \\
\addlinespace[3pt]
\multicolumn{3}{@{}l}{\textit{Strategic interaction}} \\\addlinespace[3pt]
\textbf{Beauty contest} (009)\par \citet{nagel1995unraveling} & Guesses a number from 0 to 100; the winner is closest to $p$ times the average guess ($p=1/2$, $2/3$ or $4/3$). Four rounds with feedback. & Value of $p$ and round $\rightarrow$ guesses and convergence \\\addlinespace[3pt]
\textbf{PD and Newcomb} (010)\par \citet{Shafir1992ThinkingTU} & Plays one-shot prisoner's dilemmas with the opponent's move known or unknown, faces Newcomb's two-box problem with a computer predictor, and decides whether to pay to learn the opponent's move. & Known vs.\ unknown opponent move $\rightarrow$ cooperation; paying for information \\\addlinespace[3pt]
\textbf{Bargaining} (011)\par \citet{Forsythe1994FairnessIS} & Proposes how to split \$5 or \$10 with an anonymous partner, who must accept (dictator game) or may reject (ultimatum game); pay is real or hypothetical. & Game type, pay and stake $\rightarrow$ offers \\\addlinespace[3pt]
\textbf{Trust game} (012)\par \citet{Berg1995TrustRA} & As sender, chooses how much of \$10 to send; the amount is tripled and the receiver chooses how much to return. One condition shows results from an earlier session. & Social history $\rightarrow$ amounts sent and returned \\
\addlinespace[3pt]
\multicolumn{3}{@{}l}{\textit{Social psychology}} \\\addlinespace[3pt]
\textbf{Intentional action} (005)\par \citet{article} & Reads about a company chairman (or an army lieutenant) who knowingly causes a harmful or helpful side effect, then judges whether it was intentional and how much blame or praise is deserved. & Harmful vs.\ helpful side effect $\rightarrow$ intentionality, blame and praise \\\addlinespace[3pt]
\textbf{Impression formation} (006)\par \citet{asch1946forming} & Reads a list of personality traits (e.g., intelligent, skillful, warm or cold), describes the person, and picks between paired traits such as generous/ungenerous. & Swapped, removed or reordered traits $\rightarrow$ overall impression \\\addlinespace[3pt]
\textbf{Minimal groups} (007)\par \citet{gagnon1996discrimination} & Is assigned to one of two arbitrary groups by a coin toss, then allocates course points to anonymous in-group and out-group members. & Interdependence and group identification $\rightarrow$ in-group favoritism \\\addlinespace[3pt]
\textbf{Pluralistic ignorance} (008)\par \citet{Prentice1993PluralisticIA} & Rates own comfort with campus drinking and the comfort of the average student and of friends, and answers questions about a campus keg ban. & Self vs.\ peer ratings $\rightarrow$ perceived-norm gap \\
\bottomrule
\end{tabular}
\end{table}

Two cases clarify the distinction between generated agent data and unavailable human records. In the birth-sequence task within Study~004, agents produce individual numerical answers; the evaluator counts how often those answers fall below the normative value of 72 and compares that proportion with the published human result. In Study~008, agents give individual self, peer, and friend ratings, which the evaluator combines into paired and group-level contrasts. Neither comparison requires the original human participants' response files.

\subsection{Follow-up evidence for scored behavioral hypotheses}
\label{app:finding-evidence}
Each scored behavioral hypothesis was classified by the subsequent independent human evidence for its contrast. \emph{Direct} support means that a later independent experiment or meta-analysis tests the same or a very close manipulation--outcome contrast and reports a same-direction result. \emph{Related} support means that later human evidence establishes the broader effect but does not test this specific contrast. Hypotheses with mixed or contrary follow-up results, or without located follow-up evidence, are excluded from every reported score. Table~\ref{tab:finding-evidence} gives, for each study, the number of scored hypotheses with direct and related support (45 in total) and the follow-up evidence.

\begin{table}[!htbp]
\caption{\textbf{Follow-up evidence for scored behavioral hypotheses.} Number of scored hypotheses per study with direct or related support, and the follow-up human evidence.}
\label{tab:finding-evidence}
\centering\footnotesize
\setlength{\tabcolsep}{4pt}
\renewcommand{\arraystretch}{1.1}
\begin{tabular}{@{}>{\raggedright\arraybackslash}p{0.27\linewidth}cc>{\raggedright\arraybackslash}p{0.5\linewidth}@{}}
\toprule
\textbf{Study} & \textbf{Direct} & \textbf{Related} & \textbf{Follow-up evidence} \\
\midrule
False consensus (001) & 1 & 4 & \citet{mullen1985false,klein2018manylabs2} \\
Anchoring (002) & 2 & 2 & \citet{klein2014investigating,schley2026anchoring,li2021anchoringwtp} \\
Framing (003) & 2 & 2 & \citet{klein2018manylabs2,li2025mentalaccounting,ruggeri2020prospect} \\
Representativeness (004) & 6 & 0 & \citet{mayiwar2024representativeness} \\
Intentional action (005) & 2 & 2 & \citet{klein2018manylabs2,fillon2025blame} \\
Impression formation (006) & 2 & 3 & \citet{nauts2014forming,anderson1961primacy} \\
Minimal groups (007) & 0 & 1 & \citet{schmidt2023minimal} \\
Pluralistic ignorance (008) & 1 & 2 & \citet{schroeder1998exposing} \\
Beauty contest (009) & 3 & 3 & \citet{ho1998iterated} \\
PD and Newcomb (010) & 0 & 1 & \citet{eliaz2010paying,bennett2016intrinsic} \\
Bargaining (011) & 0 & 2 & \citet{cochard2021social} \\
Trust game (012) & 2 & 2 & \citet{johnson2011trust,jiang2015financial} \\
\midrule
Total & 21 & 24 & \\
\bottomrule
\end{tabular}
\end{table}
\endgroup

\section{Complementary Experimental Results}
\label{app:Moreexp}

\subsection{Model names and OpenRouter identifiers}
\label{app:model-id}
All single-model evaluations use the OpenRouter API. Table~\ref{tab:model-openrouter} lists the display name used in this report and the exact OpenRouter model ID (provider/series) used for each run.

\begin{table*}[h]
\centering
\caption{Model display names and OpenRouter model IDs. All runs use the OpenRouter API; the ``OpenRouter model ID'' is the exact \texttt{model} string sent to the API.}
\label{tab:model-openrouter}
\begin{tabular}{@{}ll@{}}
\toprule
\textbf{Model} & \textbf{OpenRouter model ID} \\
\midrule
Claude Haiku 4.5   & \texttt{anthropic/claude-haiku-4.5} \\
DeepSeek V3.2      & \texttt{deepseek/deepseek-v3.2} \\
Gemini 3 Flash     & \texttt{google/gemini-3-flash-preview} \\
Mistral Nemo       & \texttt{mistralai/mistral-nemo} \\
Gemma 4 26b        & \texttt{google/gemma-4-26b-a4b-it} \\
GPT 5 Nano         & \texttt{openai/gpt-5-nano} \\
GPT OSS 120b       & \texttt{openai/gpt-oss-120b} \\
GPT OSS 20b        & \texttt{openai/gpt-oss-20b} \\
Qwen3 Next 80b     & \texttt{qwen/qwen3-next-80b-a3b-instruct} \\
Grok 4.1 Fast      & \texttt{x-ai/grok-4.1-fast} \\
\bottomrule
\end{tabular}
\end{table*}

\subsection{Bootstrap Standard Errors: Methodology and Justification}
\label{app:se}

\newrevision{Uncertainty for PAS (Probability Alignment Score) is quantified by resampling generated agent records. This describes uncertainty conditional on those records; no original human participant records enter the bootstrap.}

\begin{table*}[h]\small
\centering
\caption{\textbf{Total PAS and Total SE.} Each cell reports Total PAS with its bootstrap Total SE in parentheses; Total PAS is the simple mean of per-study raw PAS, and Total SE is propagated from per-study bootstrap SEs as $\sqrt{\sum \mathrm{SE}_i^2}/N$. Both are computed on the scored behavioral hypotheses; Total PAS matches Table~\ref{tab:pas-ecs-raw}.}
\label{tab:pas-pas-agg-horizontal}
\setlength{\tabcolsep}{4pt}
\renewcommand{\arraystretch}{1.0}
\begin{tabular}{@{}l cccc@{}}
\toprule
\textbf{Model} & \textbf{A1} & \textbf{A2} & \textbf{A3} & \textbf{A4} \\
\midrule
Claude Haiku 4.5 & $0.440$ {\scriptsize(0.0166)} & $0.432$ {\scriptsize(0.0159)} & $0.427$ {\scriptsize(0.0107)} & $0.394$ {\scriptsize(0.0083)} \\
DeepSeek V3.2 & $0.433$ {\scriptsize(0.0141)} & $0.428$ {\scriptsize(0.0154)} & $0.450$ {\scriptsize(0.0130)} & $0.448$ {\scriptsize(0.0144)} \\
Gemini 3 Flash & $0.542$ {\scriptsize(0.0103)} & $0.475$ {\scriptsize(0.0193)} & $0.539$ {\scriptsize(0.0086)} & $0.476$ {\scriptsize(0.0306)} \\
Mistral Nemo & $0.462$ {\scriptsize(0.0122)} & $0.428$ {\scriptsize(0.0182)} & $0.458$ {\scriptsize(0.0136)} & $0.484$ {\scriptsize(0.0189)} \\
Gemma 4 26b & $0.489$ {\scriptsize(0.0064)} & $0.473$ {\scriptsize(0.0156)} & $0.511$ {\scriptsize(0.0074)} & $0.442$ {\scriptsize(0.0094)} \\
GPT 5 Nano & $0.515$ {\scriptsize(0.0168)} & $0.501$ {\scriptsize(0.0296)} & $0.539$ {\scriptsize(0.0198)} & $0.589$ {\scriptsize(0.0157)} \\
GPT OSS 120b & $0.359$ {\scriptsize(0.0169)} & $0.397$ {\scriptsize(0.0123)} & $0.422$ {\scriptsize(0.0132)} & $0.472$ {\scriptsize(0.0116)} \\
GPT OSS 20b & $0.437$ {\scriptsize(0.0135)} & $0.406$ {\scriptsize(0.0222)} & $0.464$ {\scriptsize(0.0331)} & $0.453$ {\scriptsize(0.0160)} \\
Qwen3 Next 80b & $0.514$ {\scriptsize(0.0040)} & $0.516$ {\scriptsize(0.0063)} & $0.509$ {\scriptsize(0.0086)} & $0.518$ {\scriptsize(0.0098)} \\
Grok 4.1 Fast & $0.404$ {\scriptsize(0.0071)} & $0.372$ {\scriptsize(0.0034)} & $0.449$ {\scriptsize(0.0080)} & $0.421$ {\scriptsize(0.0085)} \\
\bottomrule
\end{tabular}
\end{table*}

\paragraph{Resampling.}
\newrevision{For each study and agent configuration, the archived procedure resamples generated agent response entries with replacement and recomputes the study-specific statistics. The reported results use $B=200$ unless stated otherwise. In nested multi-round or paired tasks, the implementation flattens responses before resampling and then groups selected entries back by agent ID. This is a response-level bootstrap, not a participant- or pair-cluster bootstrap; its SEs do not preserve within-agent dependence and should be read as descriptive uncertainty for the generated responses.}

\paragraph{Bootstrap distribution and SE.}
On each bootstrap sample we recompute the study-level PAS (or the same metric used in the main analysis). The bootstrap distribution of that metric is summarized by its standard deviation, which we take as the \emph{bootstrap standard error} for that study. Formally, $\widehat{\mathrm{SE}}_{\mathrm{boot}} = \mathrm{SD}(\hat\theta^*_1,\ldots,\hat\theta^*_B)$, where $\hat\theta^*_b$ is the statistic from the $b$-th bootstrap sample. No distributional assumption is made beyond the data-generating process implied by the empirical participant set.

\paragraph{Aggregation across studies.}
For each agent configuration, a single ``Total PAS'' is computed as the mean of per-study PAS over the $K$ studies. Under the assumption that study-level estimates are approximately independent, the SE of the mean is propagated as
\[
  \widehat{\mathrm{SE}}(\overline{\mathrm{PAS}}) = \frac{1}{K}\sqrt{\sum_{k=1}^K \widehat{\mathrm{SE}}_k^2}\,.
\]

\subsection{Extended Distributional Analysis}
\label{app:distributional_analysis}

The main text reports polarized test-level inferential agreement. Here we examine the underlying $p$-values and variability across tasks and agent configurations.

\begin{figure*}[h]
    \centering
    \includegraphics[width=\linewidth, trim={0 0 0 0}, clip]{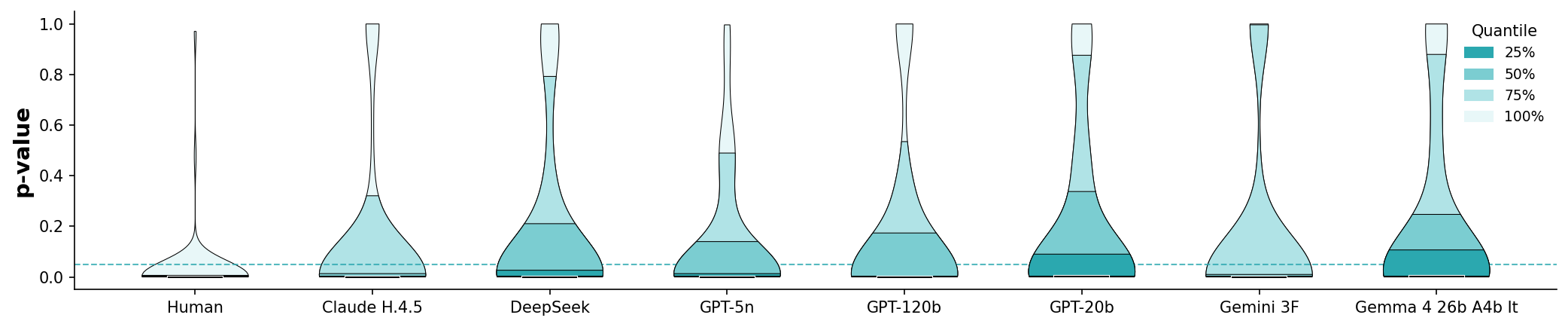}

    \caption{\textbf{Distribution of $p$-values across human and agents (A4 context).} The blue dashed line marks the standard statistical significance threshold ($p = 0.05$). The human data (far left) consistently yields highly significant results, with almost all probability mass tightly clustered near $p \approx 0$. In contrast, agent simulations fail to consistently reproduce these effects, showing a much wider spread of $p$-values with a substantial portion falling into the non-significant range ($p > 0.05$).}
    \label{fig:p_value_dist}
\end{figure*}
\begin{figure}[t]
    \centering
    \includegraphics[width=\linewidth]{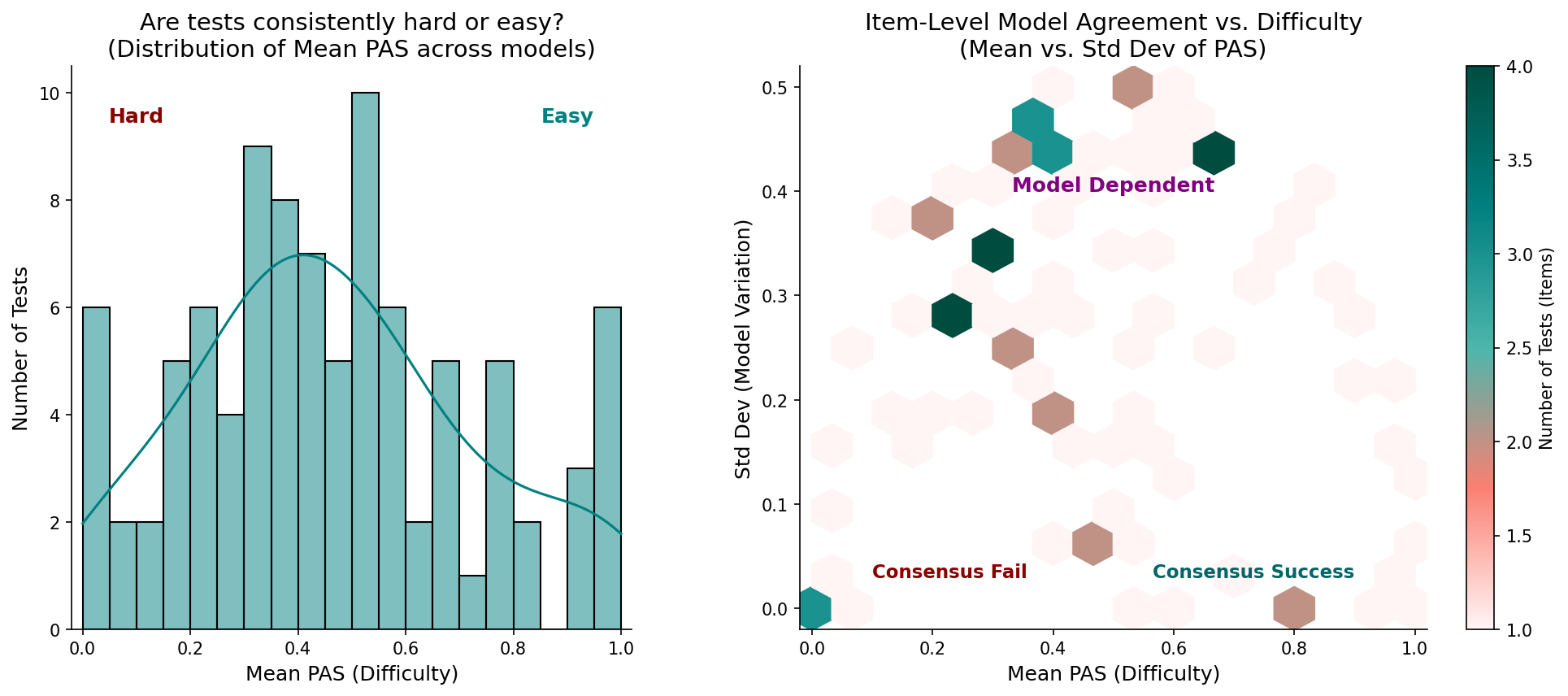} 
    \caption{\textbf{Decomposing Variance via Task Difficulty.} \textbf{(Left)} The benchmark spans a wide spectrum of difficulty levels, with variation in mean PAS. \textbf{(Right)} The Hexbin analysis identifies a ``Zone of Disagreement'' (red center), where variance peaks. Configurations differ most at intermediate average scores.}
    \label{fig:task_difficulty}
\end{figure}

\paragraph{Benchmark Sensitivity and Model Disentanglement}
To interrogate the source of variance, we decompose performance by task difficulty in Figure \ref{fig:task_difficulty}. The distribution of mean PAS (Left) demonstrates that the benchmark covers a broad difficulty spectrum rather than collapsing into a binary distribution.

The mean--variance plot shows more cross-configuration variation at intermediate mean PAS. This identifies tests on which evaluated configurations disagree, without attributing the variation to a particular architectural mechanism.

\begin{figure}[h]
    \centering
    \includegraphics[width=0.9\linewidth]{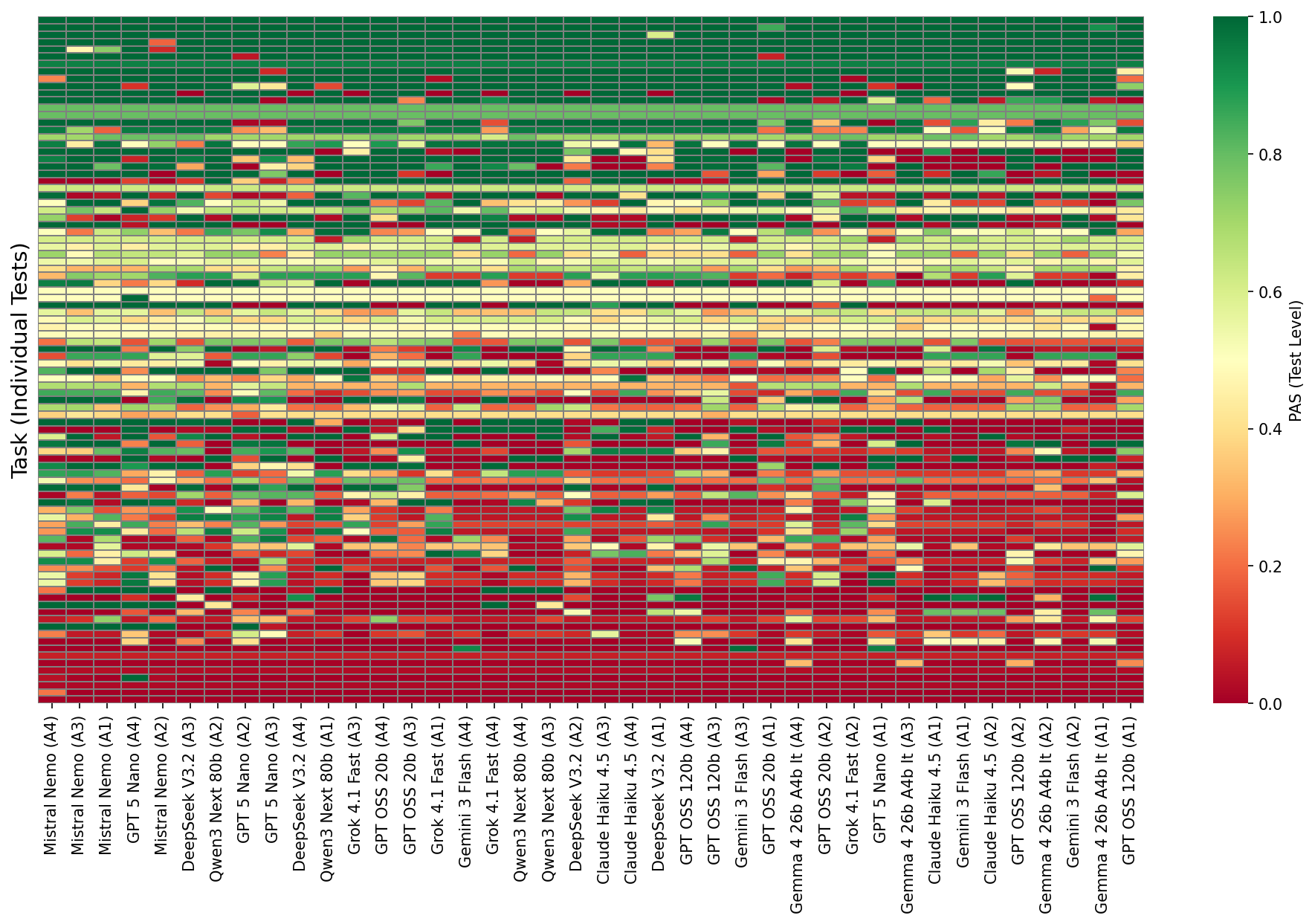}
    \caption{\textbf{The Landscape of Idiosyncratic Capabilities.} The scattered distribution of high (green) and low (red) alignment scores illustrates that capabilities are fragmented. No single agent universally dominates; instead, performance is highly specific to the interaction between agent and task type.}
    \label{fig:heatmap_matrix}
\end{figure}

\paragraph{Configuration-Specific Agreement}
Figure~\ref{fig:heatmap_matrix} shows the item-level scores across model--prompt combinations. No configuration uniformly dominates, motivating diagnosis at the effect level alongside aggregate scores.

\paragraph{Robustness to Effect-Size Heterogeneity.}
\label{app:effect_heterogeneity}
A natural concern is that the low cross-hypothesis pattern correlation $\rho$ in our RQ1 analysis of how closely agents reproduce human behavioral effects (\S\ref{sec:results}) could stem from converting heterogeneous effect-size metrics ($d$, $h$, $\eta^2$, $r$) onto a common scale. As a sensitivity check, we recompute ECS separately within each test family ($t$, $\chi^2$, $F$), avoiding any cross-family conversion; the correlation family has only three scored tests and is omitted. Averaged over agent configurations, the resulting ECS values range from $0.06$ ($F$) to $0.22$ ($t$), within the range of global ECS values in Table~\ref{tab:pas-ecs-raw}. These results suggest that the reported low concordance is not solely a consequence of pooling test families.

Figure~\ref{fig:ecs-decomposition} separates pattern correlation ($\rho$), which asks whether effects that are relatively large in the human studies are also relatively large in the simulations, from the calibration factor ($C_b$), which asks whether the two sets of effects have similar overall averages and spread. For GPT-5 Nano with role-play instructions, calibration is high ($C_b\approx0.96$) but pattern correlation is much lower ($\rho\approx0.30$). Its effects therefore occupy a broadly similar numerical range while frequently assigning the wrong relative magnitude to particular tests.

\paragraph{A Sampling-Noise Reference.}
We simulated a replication of each scored test at its original design and sample size, treating the published result as the true effect, and scored it against the human reference as for agents ($B=1000$). The mean PAS was $0.848$, below one because not all published tests are strongly significant; the corresponding ECS was $0.972$.

\subsection{Hypothesis Testing Details}
\label{app:hypothesis_details}

\begin{table}
\caption{\textbf{Summary of Hypothesis Tests.} Results based on aggregated PAS scores. Exploratory, uncorrected tests; $^{*}p<0.05$, $^{**}p<0.01$, $^{***}p<0.001$. Study-bootstrap intervals for the same design changes include zero (Table~\ref{tab:design-sensitivity}).}
\label{tab:hypothesis_summary}
\centering
\small
\begin{tabular}{llcccc}
\toprule
\textbf{ID} & \textbf{Hypothesis Description} & \textbf{Method} & \textbf{$\Delta$} & \textbf{Statistic} & \textbf{Result} \\
\midrule
\multicolumn{6}{l}{\textit{Group A: Agent Design}} \\
H1 & A3 (Demo) $>$ A1 (Empty) & Wilcoxon & $+0.017$ & $W=10$ & n.s. \\
H1b & A3 (Demo) $>$ A2 (Role-Play) & Wilcoxon & $+0.034$ & $W=3^{**}$ & Significant \\
H2 & A3 (Demo) $>$ A4 (Narrative) & Wilcoxon & $+0.007$ & $W=21$ & n.s. \\
H3 & A1 $>$ A2 (Role-Play) & Wilcoxon & $+0.017$ & $W=10$ & n.s. \\
\midrule
\multicolumn{6}{l}{\textit{Group B: Domain Specificity}} \\
H4 & Social $>$ Cognition & Pooled $t$ & $+0.034$ & $t=2.28^{*}$ & Significant \\
H5 & Social $>$ Strategic & Pooled $t$ & $+0.180$ & $t=7.52^{***}$ & Significant \\
\midrule
\multicolumn{6}{l}{\textit{Group C: Model Scaling}} \\
H6 & Proprietary $\neq$ Open-weight & Mann-Whitney & $+0.012$ & $p=1.00$ & n.s. \\
\bottomrule
\end{tabular}
\end{table}

\newrevision{Table~\ref{tab:hypothesis_summary} reports exploratory comparisons of aggregate PAS. Agent-design contrasts use two-sided Wilcoxon signed-rank tests over the ten models; $\Delta$ is the mean paired difference. Domain comparisons are descriptive because the studies and effect types differ.}

\newrevision{In the response-mixing check, each simulated participant is drawn at random from one of the ten models and the mixture is scored like a single configuration (100 draws). The mixture reached PAS $0.522$ and ECS $0.192$; one of ten single models exceeded it on each metric (mean PAS difference $-0.060$, study-bootstrap 95\% interval $-0.142$ to $0.008$).}

\subsection{Inference Cost Analysis}
\label{app:cost}
\begin{table*}[htbp]\footnotesize
\centering
\caption{\textbf{Inference Cost Analysis (USD).} Total cost to replicate the full experimental suite. For each model, \textcolor{best1}{\textbf{teal}} marks the cheapest specification and \textcolor{worst1}{\textbf{salmon}} marks the most expensive.}
\label{tab:cost-summary-horizontal-rotated}
\setlength{\tabcolsep}{2pt}
\renewcommand{\arraystretch}{0.9}
\begin{tabular}{@{}l ccccccccccc@{}}
\toprule

 & \shortstack{Claude\\H.4.5}
 & \shortstack{DeepSeek\\V3.2}
 & \shortstack{Gemini\\3 Flash}
 & \shortstack{Mistral\\Nemo}
 & \shortstack{Gemma 4\\26b}
 & \shortstack{GPT 5\\Nano}
 & \shortstack{GPT\\OSS 120b}
 & \shortstack{GPT\\OSS 20b}
 & \shortstack{Qwen3\\Next80b}
 & \shortstack{Grok 4.1\\Fast}
 & \shortstack{Mixed\\Models} \\
\midrule
A1 & 9.55 & \cellcolor{best1}{0.81} & 2.99 & 0.39 & \cellcolor{worst1}{0.62} & \cellcolor{worst1}{6.45} & 1.70 & \cellcolor{worst1}{1.42} & \cellcolor{best1}{0.82} & 0.59 & \cellcolor{worst1}{2.35} \\
A2 & \cellcolor{worst1}{10.37} & 0.81 & \cellcolor{best1}{2.82} & 0.38 & \cellcolor{best1}{0.55} & 2.83 & 1.69 & 1.38 & 0.84 & \cellcolor{best1}{0.51} & 2.15 \\
A3 & \cellcolor{best1}{6.55} & 1.07 & 2.84 & \cellcolor{worst1}{0.44} & 0.59 & 2.57 & \cellcolor{worst1}{1.74} & \cellcolor{best1}{1.18} & 0.87 & 0.85 & \cellcolor{best1}{1.65} \\
A4 & 8.37 & \cellcolor{worst1}{3.06} & \cellcolor{worst1}{5.33} & \cellcolor{best1}{0.20} & 0.61 & \cellcolor{best1}{0.40} & \cellcolor{best1}{1.01} & 1.24 & \cellcolor{worst1}{0.97} & \cellcolor{worst1}{1.29} & 1.71 \\
\bottomrule
\end{tabular}
\end{table*}

\subsection{\newrevision{Effect-specific comparisons for RQ2}}
\label{app:effect-cases}
\newrevision{The direction analysis uses each combination of a statistical test and an agent configuration for which the human posterior favors a positive or negative effect. We classify the agent posterior as favoring the reported direction, the opposite direction, or no effect. The magnitude analysis uses the subset in which the direction matches, both standardized effects are finite, and the published human effect satisfies $|d_h|\geq0.2$, the conventional lower bound for a small effect~\citep{cohen1988}. Within this subset, an absolute agent-to-human effect ratio below $0.5$ is classified as attenuation and a ratio above $2$ as exaggeration. The factor-of-two threshold brackets the average ratio of replication to original effect sizes reported by large replication projects~\citep{opensciencecollaboration2015,camerer2018replicability}. Comparisons without a directional human reference are excluded rather than counted as disagreement.}

\newrevision{The examples in Table~\ref{tab:effect-discrepancies} were checked against parsed final answers and published human summaries. In the framing task, most agent configurations favored the certain option under both wordings, rather than showing the human gain/loss reversal. For the intentional-action task, the configuration-level harm/help medians were 95.7\%/0\% versus 82\%/23\% in the published human sample. A further attenuation example is peer comfort: the median simulated self--peer gap was 0.63 points versus the published 1.71-point gap.}

\subsection{Hypothesis-level agent-design comparisons for RQ3}
\label{app:design-sensitivity}
This exploratory analysis pairs the same model and behavioral hypothesis across adjacent agent designs, so each difference measures the change associated with adding one layer of participant specification. We first average paired changes across models within each hypothesis and then balance the hypothesis-level changes across studies, so studies with more hypotheses do not dominate. The exploratory intervals resample studies.

\begin{table}[h]
\footnotesize\centering
\caption{\textbf{Hypothesis-level changes in PAS.} Mean paired differences and study-bootstrap 95\% intervals. Hypothesis counts describe the ten-model mean difference: increase $>0.02$, decrease $<-0.02$, or within $\pm0.02$.}
\label{tab:design-sensitivity}
\begin{tabular}{@{}lccccc@{}}
\toprule
\textbf{Change} & \textbf{Mean $\Delta$PAS} & \textbf{95\% interval} & \textbf{Increase} & \textbf{Decrease} & \textbf{Within $\pm0.02$} \\
\midrule
A1 $\rightarrow$ A2 & $-0.017$ & $[-0.049,\ 0.012]$ & 15 & 15 & 15 \\
A2 $\rightarrow$ A3 & $+0.034$ & $[-0.008,\ 0.076]$ & 21 & 12 & 12 \\
A3 $\rightarrow$ A4 & $-0.007$ & $[-0.059,\ 0.032]$ & 15 & 11 & 19 \\
\bottomrule
\end{tabular}
\end{table}

The effect of added context is model-specific: relative to A1, backstories (A4) raised PAS for GPT OSS 120b and GPT-5 Nano but lowered it for Claude Haiku 4.5, Gemini 3 Flash, and Gemma 4 26b, each by more than twice the response-bootstrap standard error of the difference.

\subsection{Prompt Sensitivity Ablation}
\label{app:prompt-sensitivity}

We assessed sensitivity to prompt wording in the A3 (Demographic) specification. We created two paraphrased variants by rewriting all three components---role-playing framing, demographic formatting, and task instructions---while preserving informational content (Table~\ref{tab:prompt-sensitivity-design}). We measured within-paraphrase PAS variation (max--min range across the three A3 variants) on Study~2 ($N=256$) and Study~8 ($N=468$), selected for their moderate sample sizes, and compared it with the A1--A4 range from the same rerun (Table~\ref{tab:prompt-sensitivity-results}).

\begin{table}[!h]\small
\centering
\caption{\textbf{Prompt paraphrase design.} Three A3 variants with equivalent information but different surface forms.}
\label{tab:prompt-sensitivity-design}
\setlength{\tabcolsep}{3pt}
\begin{tabular}{@{}lccc@{}}
\toprule
 & \textbf{Role frame} & \textbf{Demographics} & \textbf{Task instructions} \\
\midrule
A3 & ``You are participating\ldots'' & Bullet list & ``Follow the experimenter's\ldots'' \\
Para.\ A & ``Imagine you are a person\ldots'' & Inline & ``Respond to each task\ldots'' \\
Para.\ B & ``You have been recruited\ldots'' & Indented block & ``Answer every question\ldots'' \\
\bottomrule
\end{tabular}
\end{table}

\begin{table}[!h]\small
\centering
\caption{\textbf{Prompt sensitivity results.} PAS range (percentage points) across the three A3 wordings and across the four agent designs, computed on the scored behavioral hypotheses.}
\label{tab:prompt-sensitivity-results}
\begin{tabular}{@{}llcc@{}}
\toprule
\textbf{Model} & \textbf{Study} & \textbf{Paraphrase range} & \textbf{A1--A4 range} \\
\midrule
Mistral Nemo & Study 2 ($N{=}256$) & 2.4 & 3.2 \\
Mistral Nemo & Study 8 ($N{=}468$) & 6.0 & 13.0 \\
DeepSeek V3.2 & Study 2 ($N{=}256$) & 1.3 & 8.1 \\
DeepSeek V3.2 & Study 8 ($N{=}468$) & 13.3 & 24.4 \\
\bottomrule
\end{tabular}
\end{table}

Across the four tested model--study combinations, PAS variation from paraphrasing stays within 13.3 percentage points and is smaller than the A1--A4 range in every combination (3.2 to 24.4 percentage points).

\section{Broader Impacts}
\phantomsection\label{sec:impact}%
\subsection{Potential Benefits}
\label{community}\label{commnity}

\textsc{HumanStudy-Bench} makes published experimental protocols reusable for evaluating AI agents. Researchers can use it to test whether an agent reproduces a specific behavioral finding, identify where its responses differ from human evidence, and examine how those differences change with model or agent design. The platform may also reduce the repeated effort required to reconstruct experimental materials, conditions, and analyses. By evaluating population-level effects rather than surface-level fluency, it provides evidence that is more diagnostic than a single overall judgment of whether an agent appears human-like. The benchmark is available for research use and is designed to support contributions of additional studies and human reference evidence.

\subsection{Risks and Responsible Use}
\label{ethic}

\textbf{Limits of interpretation and use.}
Agreement on the included findings concerns particular effects under particular experimental conditions. Our results show that agents can miss or reverse a human effect, preserve its direction while distorting its magnitude, and change unevenly with agent design. Benchmark scores should therefore not be used on their own for consequential decisions in areas such as public policy, health, hiring, education, or market research; new uses require validation against relevant human data.

\textbf{Representativeness and stereotypes.}
The benchmark draws on 12 original studies published between 1946 and 2003 (Table~\ref{tab:hypothesis-teaser} shows examples). Their samples are predominantly WEIRD (Western, educated, industrialized, rich, and democratic)---mainly US, UK, Canadian, and Dutch university students, with one US public-park sample \citep{article} as the only non-student exception. The follow-up evidence extends to 2026 and includes some cross-national evidence, notably a 19-country replication \citep{ruggeri2020prospect}, but is still drawn largely from Western, educated populations. Findings may also change over time. In addition, demographic attributes and backstories may lead models to rely on stereotypes rather than represent the people described. Benchmark agreement should therefore not be interpreted as universal human agreement, and representational harms should be evaluated separately from PAS and ECS.

\textbf{Materials and reproducibility.}
The benchmark uses materials derived from published, peer-reviewed studies for non-commercial research. We recruited no new human participants and collect no personally identifiable information. Released materials should follow applicable licenses and publisher restrictions, with sensitive or deceptive materials documented in context. Reconstructed protocols and study specifications undergo human review, and the statistical analyses are deterministic for fixed inputs. Model responses, however, can vary with sampling, model updates, and provider-side changes; results should therefore report the model version, agent configuration, prompts, sampling settings, and execution date.

\end{document}